\documentclass[12pt,useAMS,usenatbib,nomarkers,referee]{biom}

\usepackage[T1]{fontenc}
\usepackage[utf8]{inputenc}
\usepackage{lmodern}
\usepackage{textcomp}

\usepackage{fancyhdr}
\usepackage{amsmath,amssymb,bm,dsfont,mathrsfs}
\usepackage[mathscr]{euscript}
\allowdisplaybreaks

\usepackage{thmtools,thm-restate}

\usepackage{graphicx}
\graphicspath{{./figs/}}
\usepackage{xcolor}
\usepackage{float}
\usepackage{subcaption}   

\usepackage{setspace}
\usepackage{enumitem}
\usepackage{titlesec}
\titlespacing*{\section}{0pt}{.25\baselineskip}{.1\baselineskip}
\titlespacing*{\subsection}{0pt}{.25\baselineskip}{.1\baselineskip}
\titlespacing*{\subsubsection}{0pt}{0.25em}{0.25em}

\usepackage{booktabs}
\usepackage{multirow}
\usepackage{tabularx}
\usepackage{threeparttable}
\usepackage{pdflscape}

\makeatletter
\@ifundefined{thead}{}{}
\makeatother
\usepackage{makecell}

\usepackage{siunitx}
\newcolumntype{L}{>{\raggedright\arraybackslash}X}
\newcolumntype{C}{>{\centering\arraybackslash}m{0.9cm}} 
\newcolumntype{A}{S[table-format=1.3]} 
\newcolumntype{O}{S[table-format=1.3]} 
\newcolumntype{R}{S[table-format=1.2]} 

\usepackage{algorithm}
\usepackage{algpseudocode}

\usepackage{authblk}
\usepackage{multicol}
\usepackage{comment}
\usepackage[normalem]{ulem} 
\usepackage{url}

\usepackage[hidelinks]{hyperref}

\newcommand{\bs}{\boldsymbol}

\usepackage{orcidlink}

\title{Robust Spectral Fuzzy Clustering of Multivariate Time Series with
Applications to Electroencephalogram}

\author{
Ziling Ma\,\orcidlink{0009-0006-0323-6467}\textsuperscript{1,*}, 
Mara Sherlin Talento\,\orcidlink{0000-0003-2375-8634}\textsuperscript{1,*}, 
Ying Sun\,\orcidlink{0000-0001-6703-4270}\textsuperscript{1}, 
Hernando Ombao\,\orcidlink{0000-0001-7020-8091}\textsuperscript{1}
}

    \author{
Ziling Ma\,\orcidlink{0009-0006-0323-6467}\textsuperscript{1,*}, 
Mara Sherlin Talento\,\orcidlink{0000-0003-2375-8634}\textsuperscript{1,*}, 
Ying Sun\,\orcidlink{0000-0001-6703-4270}\textsuperscript{1}, 
Hernando Ombao\,\orcidlink{0000-0001-7020-8091}\textsuperscript{1}
\thanks{King Abdullah University of Science and Technology (KAUST), Computer, Electrical and Mathematical Sciences and Engineering (CEMSE) Division, Thuwal 23955-6900, Saudi Arabia.}
\thanks{*These authors contributed equally to this work.}
\thanks{Correspondence: ziling.ma@kaust.edu.sa, marasherlin.talento@kaust.edu.sa, ying.sun@kaust.edu.sa, hernando.ombao@kaust.edu.sa}
\thanks{Preprint. Under review.}
}

\begin{document}

   \pagerange{\pageref{firstpage}--\pageref{lastpage}} \pubyear{2024}

\label{firstpage}


\begin{abstract}

Clustering multivariate time series (MTS) is challenging due to non-stationary cross-dependencies, noise contamination, and gradual or overlapping state boundaries. We introduce a robust fuzzy clustering framework in the spectral domain that leverages Kendall’s tau–based canonical coherence to extract frequency-specific monotonic relationships across variables. 
Our method takes advantage of dominant frequency-based cross-regional connectivity patterns to improve clustering accuracy while remaining resilient to outliers, making the approach broadly applicable to noisy, high-dimensional MTS.
Each series is projected onto vectors generated from a spectral matrix specifically tailored to capture the underlying fuzzy partitions. 
Numerical experiments demonstrate the superiority of our framework over existing methods. As a flagship application, we analyze electroencephalogram recordings, where our approach uncovers frequency- and connectivity-specific markers of latent cognitive states such as alertness and drowsiness, revealing discriminative patterns and ambiguous transitions. 
\end{abstract}

\vspace{1em}

\maketitle

\section{Introduction}

Brain networks, composed of nodes representing brain regions and edges representing cross-regional connectivity, are central to understanding neural dynamics and distinguishing between cognitive states and populations \citep{wu2025wavelet}. Clustering multivariate time series (MTS) of brain signals enables the discovery of latent neural patterns, detection of cognitive transitions, and development of personalized neuroscience applications \citep{honey2007network,van2010exploring}. Electroencephalogram (EEG) signals are particularly challenging: they are high-dimensional, temporally dependent, spectrally complex \citep{von2007tutorial}, and contaminated by artifacts such as eye blinks or muscle activity \citep{rodrigues2017noise,jiang2019removalartifacts}. Moreover, brain states (such as alertness and drowsiness) often evolve gradually with overlapping boundaries \citep{masulli2019fuzzy,zhao2024leveraging}, making fuzzy clustering more appropriate than crisp partitioning.

Spectral-domain analysis provides a principled way to summarize oscillatory interactions and capture functional connectivity across regions \citep{van2007comparison,ombao2024spectral}. EEG oscillations in specific bands have been linked to neurological disorders \citep{Little2014BetaParkinson} and task-related attention \citep{Li2022exploratoryDrivers}, underscoring the discriminative power of frequency-specific features. Thus, robust fuzzy clustering methods that incorporate spectral and spatial information are highly desirable for EEG analysis.

However, existing fuzzy clustering approaches for MTS remain limited. Most operate in the time domain, relying on temporal features, autoregressive models, or correlation structures \citep{izakian2015fuzzy,ann2010wavelet}. Robustness has been considered through trimming strategies \citep{d2023robust} or quantile cross-spectral features \citep{lopez2022quantile}, but these are computationally intensive. Subspace-based methods such as \citet{ma2025fcpca} improve scalability but still neglect cross-regional connectivity. Overall, current methods are either time-domain only, sensitive to contamination, or fail to leverage spatial connectivity—leaving a gap for a robust, frequency-domain fuzzy clustering framework that can handle overlapping EEG states.

We propose FuzzCoh, a fuzzy clustering framework in the spectral domain that leverages Kendall’s tau–based canonical coherence (KenCoh) \citep{talento2024kencoh} to extract robust connectivity features. FuzzCoh allows partial membership assignments, capturing transitional or mixed-state phenomena, and provides interpretable insights into frequency-specific and connectivity-specific dependencies. We validate FuzzCoh on multi-node, multi-trial EEG data for detecting drowsy versus alert states \citep{Cao2019multi}, a problem with significant safety implications \citep{tefft2010asleep,higgins2017asleep,filtness2017sleep}. Since drowsiness and alertness are not strictly binary but instead exhibit fuzzy transitional phases \citep{albadawi2022review,ma2025fcpca}, fuzzy clustering is particularly well suited.\footnote{\scriptsize To demonstrate that FuzzCoh is not limited to EEG data, we provide an additional application on motion MTS data in the Appendix, Section \ref{cricket_data}.}
Our contributions are threefold: (1) we develop a novel frequency-domain fuzzy clustering method, FuzzCoh, which identifies oscillatory bands that discriminate between different MTS states; (2) we incorporate robust estimation of spectral dependence to maintain stability under contamination; and (3) by examining cross-regional connectivity, we account for spatial information and link brain states to functional networks, thereby providing deeper neuroscientific insights.

\section{Methodology} \label{sec:methods}

\subsection{Lagged dependence matrix of filtered signals}
Let ${\bs{Z}_{BT\times(p+q)} = (\bs{X}_{BT\times p}, \bs{Y}_{BT\times q})}$ be a non-stationary MTS. Suppose that this process is `locally stationary' so that for a block $b, (b = 1, \dots , B)$ of length $T$, the time series are approximately stationary. Denote ${\bs{Z}^{(b)}_{\Omega}(t) \in \mathbb{R}^{p+q}}$ as the $\bs{Z}_t$ corresponding to the $b$-th block filtered to the $\Omega$ frequency band.
Furthermore, let 
\begin{gather}
    \zeta_{0,\Omega}^{(b)} = Z^{(b)}_{\Omega}(t) - Z^{(b)}_{\Omega}(s), \notag \\
    \zeta_{\ell,\Omega}^{(b)} = Z^{(b)}_{\Omega}(t-\ell) - Z^{(b)}_{\Omega}(s-\ell), \notag \\
    {\tau}^{(b)}(\Omega, \ell) = \mathbb{E}[\text{sign}\{ \zeta_{0,\Omega}^{(b)} \zeta_{\ell,\Omega}^{(b)} \}], \label{ktau} \notag \\
     f(Z_{\Omega,j}(t-\ell), Z_{\Omega,k}(t)) := \sin\left(\frac{\pi}{2}{\tau}^{(b)}(\Omega, \ell)\right). \label{tsksm}
\end{gather}
for $t \neq s$ and a fixed $\ell$, i.e., $f: \mathbb{R}^{2} \rightarrow (-1,1)$.
\cite{talento2024kencoh} define the $b$-th block lagged dependence matrix among filtered signals on $\Omega$ at lag ${\ell = 0, \pm 1, \dots, \pm L}$ as
\begin{equation}
    \bs{P}^{(b)}(\Omega, \ell) = \begin{pmatrix}
    \bs{P}^{(b)}_{X,X}(\Omega, \ell) & \bs{P}^{(b)}_{X,Y}(\Omega, \ell) \\
    \bs{P}^{(b)}_{Y,X}(\Omega, \ell) & \bs{P}^{(b)}_{Y,Y}(\Omega, \ell)
\end{pmatrix}, \label{Pmat}
\end{equation}
\begin{align*}
   \text{where} \ {P}^{(b)}_{jk}(\Omega, \ell) &:= f(Z_{\Omega,j}(t), Z_{\Omega,k}(t+\ell)) \\
    &= f(Z_{\Omega,j}(t-\ell), Z_{\Omega,k}(t))
\end{align*}
for $j,k = 1, \dots, p+q$.
We then extract the cross-vector topologies at block $b$ and frequency-band $\Omega$, denoted as $\bs{u}_{b,\Omega}$ and $\bs{v}_{b,\Omega}$, through the following objective function in \cite{talento2024kencoh}, and denoted as $g^{\Omega}(b)$, i.e.,
\begin{gather*} 
    g^{\Omega}(b) := \max_{\bs{u}_{b,\Omega},\bs{v}_{b,\Omega}, \ell} \left\{\bs{u}^\top_{b,\Omega}\bs{P}^{(b)}_{X,Y}(\Omega,\ell)\bs{v}_{b,\Omega}\right\}^2, \ \text{such that} \\
    \bs{u}^\top_{b,\Omega}\bs{P}^{(b)}_{X,X}(\Omega,0)\bs{u}_{b,\Omega} = \bs{v}^\top_{b,\Omega}\bs{P}^{(b)}_{Y,Y}(\Omega,0)\bs{v}_{b,\Omega} = 1. 
\end{gather*}
These $\bs{u}_{b,\Omega}$ and $\bs{v}_{b,\Omega}$, called the block-wise canonical variates \citep[see][for details]{talento2024kencoh}, show the contribution of the variables relative to $g^{\Omega}(b)$. Thus, this paper exploited the connectivity-structure through $\bs{u}_{b,\Omega}$ and $\bs{v}_{b,\Omega}$, and determine the clusters of multiple time series for a specific oscillation band. 

\subsection{FuzzCoh: The algorithm} 
\subsubsection{The clustering procedures}
We use the standardized canonical directions, $\bs{u}_{b,\Omega}$ and $\bs{v}_{b,\Omega}$, which are computed from the filtered signals $\bs{X}^{(b)}_{\Omega} \in \mathbb{R}^{T \times p}$ and $\bs{Y}^{(b)}_{\Omega} \in \mathbb{R}^{T \times q}$, and capture the maximum cross-dependence at a given spectral band $\Omega$, as spectral features to conduct the fuzzy clustering. 

Consider a filtered MTS dataset consisting of $B$ realizations, denoted by $\bs{Z}_{\Omega} = \{ \bs{Z}^{(1)}_{\Omega}, \dots, \bs{Z}^{(B)}_{\Omega} \}$, where 
\[
\bs{Z}^{(b)}_{\Omega} = \left( \bs{X}^{(b)}_{\Omega}, \bs{Y}^{(b)}_{\Omega} \right),
\]
for $b=1, \dots, B$.
Denote ${\bs{d}_b^{\Omega} = ( \lvert\bs{u}_{b,\Omega}^\top\rvert, \lvert\bs{v}_{b,\Omega}^\top \rvert)^\top \in \mathbb{R}^{p+q},}$ 
where \(\lvert\cdot\rvert\) is the absolute-value operator, applied element-wise to a vector. It concatenates the element-wise absolute values of the standardized canonical directions obtained at frequency band $\Omega$ and $b$-th series. We compute this for each MTS object to obtain a collection of $\{\bs{d}_b^{\Omega}\}_{b = 1}^B$, denoted as $\bs{d}^{\Omega} = \{\bs{d}_1^{\Omega},\cdots,\bs{d}_B^{\Omega}\}$.
Then we perform a fuzzy $C$-means clustering model using $\bs{d}^\Omega$, aiming to find a set of cluster centers 
$\overline{\bs{d}}^{\Omega}= \left\{\overline{\bs{d}}_{1}^{\Omega}, \dots, \overline{\bs{d}}_{C}^{\Omega}\right\}$,
and the $B\times C$ fuzzy coefficient matrix, $\bs{E}=(e_{bc}^{\Omega})$, for ${b=\{1,\cdots,B\}}$ and $c=\{1,\cdots,C\}$, by solving the minimization problem
\begin{equation}\label{fuzzy_opt}
    \left\{
\begin{aligned}
&\min_{\overline{\bs{d}}^{\Omega}, \bs{E}} \sum_{b=1}^B \sum_{c=1}^C (e_{bc}^{\Omega})^m \left\| \bs{d}_b^{\Omega} - \overline{\bs{d}}_{c}^{\Omega} \right\|^2, \\
&\text{subject to} \quad \sum_{c=1}^C e_{bc}^{\Omega} = 1, \quad e_{bc}^{\Omega} \ge 0, \quad \forall b=1,\dots,B.
\end{aligned}
\right.
\end{equation}
Here, $e_{bc}^{\Omega} \in [0,1]$ represents the membership degree of the $b$-th series to the $c$-th cluster, and $m>1$ is the fuzziness parameter controlling the degree of fuzziness of the partition.

The optimization problem in Equation \ref{fuzzy_opt} is solved via an iterative procedure. The update formula for the membership takes the form
\begin{equation}
    e_{bc}^{\Omega} = \left( \sum_{c'=1}^C \left( \frac{ \left\| \bs{d}_b^{\Omega} - \overline{\bs{d}}_{(c)}^{\Omega} \right\|^2 }{ \left\| \bs{d}_b^{\Omega} - \overline{\bs{d}}_{(c')}^{\Omega} \right\|^2 } \right)^{\frac{1}{m-1}} \right)^{-1},
\end{equation}
and for the centers, we have
\begin{equation}
   \overline{\bs{d}}_{c}^\Omega = \frac{ \sum_{b=1}^B (e_{bc}^{\Omega})^m \bs{d}_b^\Omega }{ \sum_{b=1}^B (e_{bc}^{\Omega})^m }.
\end{equation}

\subsubsection{Cluster validity index for $C$ and $m$}
As stated by \cite{rhee1996validity}, if fuzzy cluster analysis is to make a significant contribution to engineering applications, greater attention must be given to the fundamental decision of determining the number of clusters in the data. In this paper, we present a cluster validity index to help determine the optimal selection of $C$ and $m$. We adopt the Fuzzy Silhouette index (FSI) \citep{rawashdeh2012crisp}, which generalizes the classical silhouette index \citep{rousseeuw1987silhouettes} to the fuzzy setting. For each MTS
object \( d_b^\Omega \), the silhouette value with respect to cluster \( c \) is computed as
\begin{equation}
s_{bc}^\Omega = \frac{n^\Omega_{bc} - a^\Omega_{bc}}{\max\left(a^\Omega_{bc}, n^\Omega_{bc}\right)},
\end{equation}

where $a^\Omega_{bc}$ is the average dissimilarity between $\bm{d}_b^\Omega$ and all other objects in cluster $c$, 
weighted by their membership degrees, defined as
\begin{equation}
a^\Omega_{bc} = 
\frac{\sum_{\substack{j=1 \\ j \neq b}}^{B} (e_{jc}^\Omega)^m \cdot \lVert \bm{d}_b^\Omega - \bm{d}_j^\Omega \rVert}
     {\sum_{\substack{j=1 \\ j \neq b}}^{B} (e_{jc}^\Omega)^m},
\end{equation}
and $n^\Omega_{bc}$ is the minimum average dissimilarity between $\bm{d}_b^\Omega$ and all objects in another cluster $c' \ne c$, defined as
\begin{equation}
n^\Omega_{bc} = 
\min_{c' \ne c} 
\left( 
\frac{\sum_{\substack{j=1 \\ j \neq b}}^{B} (e_{jc'}^\Omega)^m \cdot \lVert \bm{d}_b^\Omega - \bm{d}_j^\Omega \rVert}
     {\sum_{\substack{j=1 \\ j \neq b}}^{B} (e_{jc'}^\Omega)^m}
\right).
\end{equation}

The overall FSI, for a specific frequency band $\Omega$, is then given by
\begin{equation}
\mathrm{FSI}_\Omega = \frac{1}{B} \sum_{b=1}^{B} \sum_{c=1}^{C} (e_{bc}^{\Omega})^m \cdot s^\Omega_{bc}.
\end{equation}

\begin{figure}[htbp]
    \centering
    \includegraphics[width=1\linewidth]{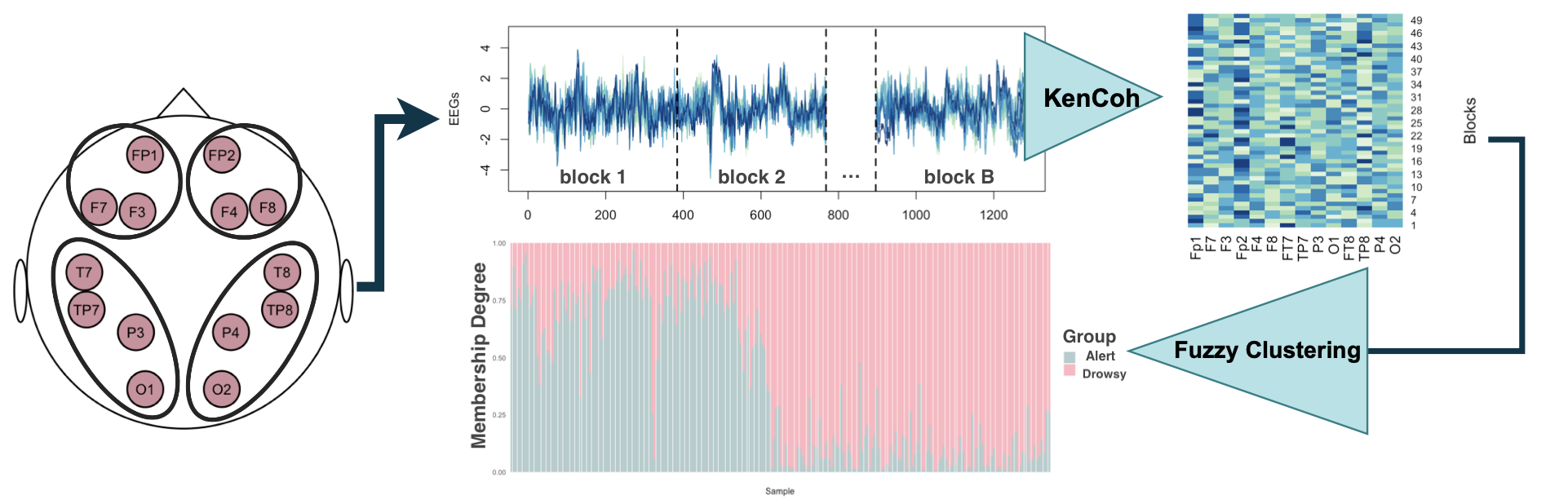}
    \caption{The FuzzCoh framework.}
    \label{framework}
\end{figure}

A higher FSI$_\Omega$ value indicates that the fuzzy memberships align well with the underlying cluster structure, reflecting compact and well-separated clusters with minimal ambiguity in membership assignments. Therefore, when no prior knowledge is available for a given dataset, we recommend performing a grid search. For example, $C\in \{2,3,\cdots,6\}$, as suggested in \cite{ferraro2019fclust,ma2025fcpca}, and $m\in [1.2, 2.5]$  (a common range for the fuzziness parameter), and selecting the configuration that yields the highest FSI. However, as in our simulation and real data application, we focus on detecting two different brain states, i.e., drowsy and alert. Thus, we fix $C=2$ in our paper. Figure \ref{framework} shows the flow of the FuzzCoh algorithm.

\section{Simulated EEG performance}
\label{sec4:simulation}

This section evaluates the performance of FuzzCoh on both clean and contaminated MTS using the simulation designs of \citet{talento2024kencoh}. Our aim is to quantify accuracy on clean data and robustness under contamination, while benchmarking against representative alternatives.

\subsection{Simulation design and settings}
We begin with a mixture of second-order autoregressive processes (AR(2)) for different blocks, that is, for eight channels/time series ($p=4$ and $q=4$) we have
\begin{gather}
    \bs{A}^{(b)}(t) = \bs{A}_1D^{(b)}(t) + \bs{A}_0(1-D^{(b)}(t)), \notag \\
    \begin{pmatrix}
        \bs{X}^{(b)}(t) \\
        \bs{Y}^{(b)}(t) \\
    \end{pmatrix} = \bs{A}^{(b)}(t)  \bs{O}^{(b)}(t) + \bs{W}^{(b)}(t)     \label{rawsim}
\end{gather}
where $\bs{X}^{(b)}(t) \in \mathbb{R}^p$ and $\bs{Y}^{(b)}(t) \in \mathbb{R}^q$ are the concatenated MTS at block $b$. The vector of latent processes, $\bs{O}^{(b)}(t) \in \mathbb{R}^5$, is composed of different mutually exclusive AR(2)'s. The parameters of ${O}_r^{(b)}(t)$, for $r = 1, \dots, 5$, are $\phi_{1,r} = 2\exp(-M)cos(2\pi \omega_r) \ \text{ and } \ \phi_2 = -\exp(-2M)$, where $M = 1.05$, $S\omega_r \in \{2,6,10,20,40\}$ and $S = 128$ Hz. Observe that $S\omega_r$ belongs to any of the five frequency bands defined in Section~\ref{sec:methods} \citep[details can be found in][]{ombao2024spectral}. 

The fuzzy series are generated through binary indicator, $D^{(b)}(t) = c$, for $c \in \{0,1\}$. That is, we generate MTS with mixing-matrix $\bs{A}_c \in \mathbb{R}^{(p+q) \times 5}$ if time point, $t$, belongs to cluster $c$. This formulation allows us to create blocks of MTS that exclusively belong to one of the two clusters, as well as blocks that contain contributions from both processes, i.e., representing fuzzy series. The $b$-th MTS is said to be pure if $D^{(b)}(1) = D^{(b)}(2) = \dots = D^{(b)}(T) = c$ and $c \in \{0,1\}$. The $b$-th MTS is said to be fuzzy if $ \mathbb{P}[D^{(b)}(t) = 0] \in (0,1)$ and $\mathbb{P}[D^{(b)}(t) = 1] = 1 - \mathbb{P}[D^{(b)}(t) = 0]$ for all $t = 1, \dots, T$.

We provide here four examples that are different in terms of the independent white-noise processes, $\bs{W}(t) \in \mathbb{R}^{p+q}$. We let the marginal distribution of white noise processes to have the following distributions, namely, (i) Example 1: $W^{(b)}_{p}(t) \sim N(0,1)$; (ii) Example 2: $W^{(b)}_{p}(t) \sim \text{Student's }t_3$; and (iii) Example 3: $W^{(b)}_{p}(t) \sim \text{Student's }t_1$. Examples in (ii) and (iii) are known to be heavy-tailed and example in (iii) has non-existence of second moment. 
In this paper, a single block consists of 384 time points, i.e., 128 Hz sampling rate for 3 seconds. Moreover, we have $B = 300$ yielding $115,200 \times 8$ data points. The runtime for the whole framework in Figure~\ref{framework} with this amount of data takes about 2 minutes for a desktop with 3 gigahertz 6-Core Intel Core i5 processor and 16 gigabytes 2667 MHz DDR4 memory. The codes are available in the Supplementary Material and implemented in parallel-run.

\subsection{Comparison methods and evaluation metrics}
To evaluate the performance of the proposed method, we benchmark it against several alternatives.\footnote{\scriptsize Code for the first three methods is in
the R package \texttt{mlmts}; the FCPCA implementation is at
\url{https://github.com/arbitraryma/FCPCA}.}
\begin{enumerate}[leftmargin=*, itemsep=0.25em, topsep=0.25em]
    \item Variable-based principal component analysis (VPCA) clustering. It reduces MTS data by applying PCA across the variable dimension, yielding a compact set of variable‑wise components that preserve spatial correlations for efficient downstream clustering \citep{he2018unsupervised}.
    \item Quantile cross-spectral density based fuzzy clustering (QCD). It captures complex dependence structures robustly, even under heavy-tailed distributions \citep{lopez2022quantile}.
    \item Wavelet-based clustering of MTS (WAV). It  decomposes each MTS into wavelet variances and wavelet correlations across multiple scales, capturing both variability within individual series and interactions between components \citep{d2012wavelets}.
    \item Fuzzy clustering of MTS using common principal component analysis (FCPCA). estimating cross-covariance matrices within each cluster, extracting common principal components, and projecting series into low-dimensional subspaces for membership updates \citep{ma2025fcpca}.
\end{enumerate}

As we work with fuzzy partitions, a membership threshold is required to assign each MTS to a specific cluster \citep{lopez2022quantile,ma2025fcpca}.  
Because the experiment distinguishes only between the {drowsy} and {alert} states, we fix the number of clusters at \(C = 2\). Following a standard convention, the \(b\)-th series (block) is assigned to cluster \(c\) whenever its membership exceeds the cut‑off,
$e_{bc}^{\Omega} > 0.7$.   For the generated {switching} series, designed to retain certain membership in both clusters, we deem the allocation correct when $\max\{e_{b1}^{\Omega},\,e_{b2}^{\Omega}\} < 0.7$, thereby recognizing their intrinsically fuzzy status \citep{maharaj2011fuzzy,d2012wavelets}.

After assigning each block a clear label, we use the Rand index (RI) to compare the results of different methods. The RI measures the concordance between our clustering partitions and the true labels. RI has range $[0, 1]$. Higher RI values indicate greater similarity between the experimental and true partitions. This setup allows us to evaluate our proposed clustering method under both well-separated and partially overlapping time-series dynamics. 

\begin{figure}[ht]
    \centering
    \includegraphics[width=0.8\linewidth]{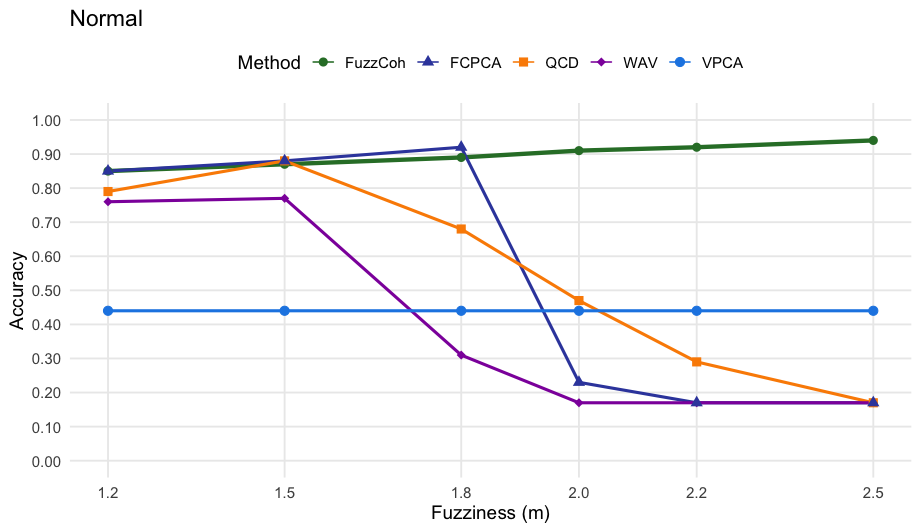}
    \caption{Simulation result under Example 1.}
    \label{normal}
\end{figure}

\begin{figure}[ht]
    \centering
    \includegraphics[width=0.8\linewidth]{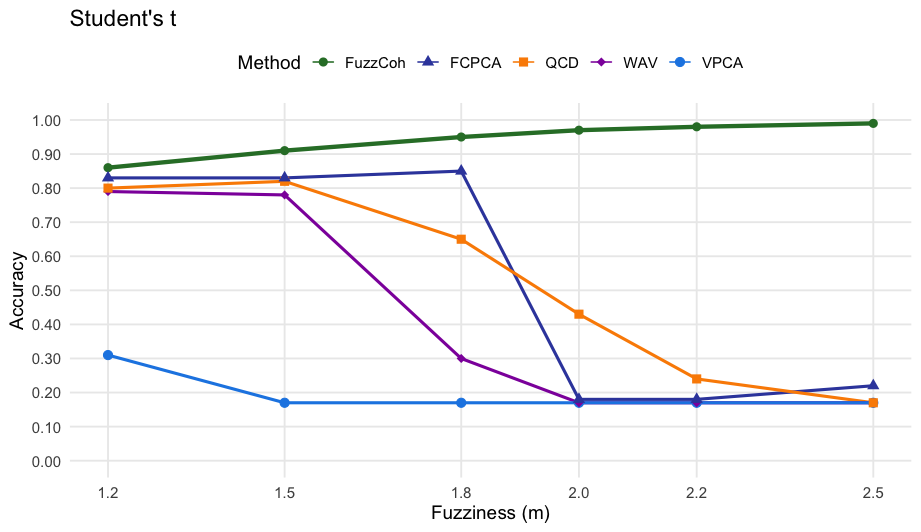}
    \caption{Simulation result under Example 2.}
    \label{student t}
\end{figure}

\begin{figure}[ht]
    \centering
    \includegraphics[width=0.8\linewidth]{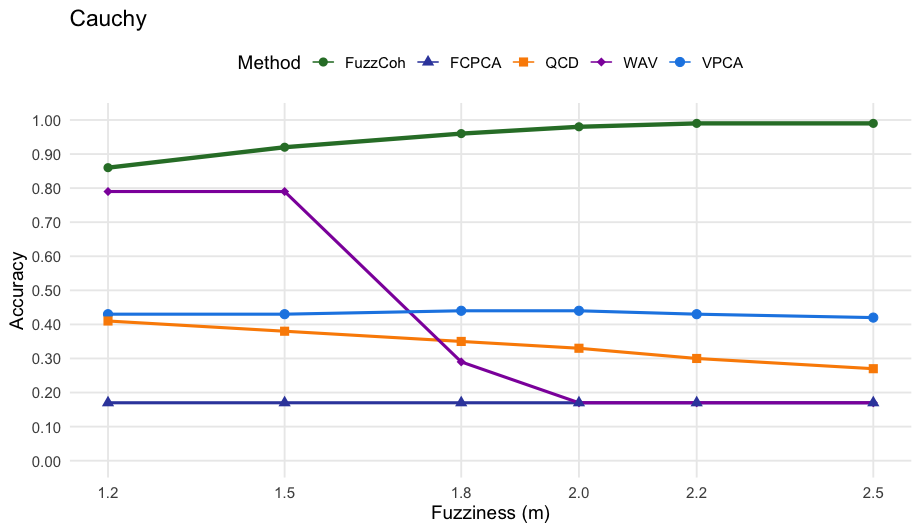}
    \caption{Simulation result under Example 3.}
    \label{cauchy}
\end{figure}

The results are shown in Figure \ref{normal}, \ref{student t}, and \ref{cauchy}, where we present the mean Rand Index (RI) of all methods across 100 replications on the same dataset, using different fuzziness parameters. To have a fair comparison, we apply all methods to the raw MTS without feature extraction from specific frequency bands. Across all scenarios, FuzzCoh consistently outperforms the competing methods. It maintains high accuracy under Normal and Student’s t settings and is uniquely robust under the most contaminated case (Cauchy), where all other methods deteriorate severely. This demonstrates that FuzzCoh is stable across different noise regimes and resilient to heavy-tailed contamination.

\begin{figure}[ht]
    \centering
    \includegraphics[width=0.8\linewidth]{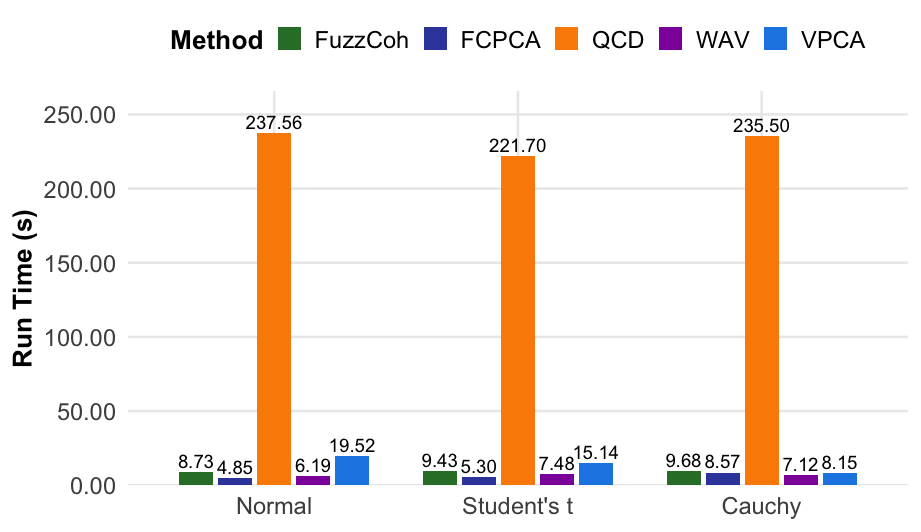}
    \caption{Average times comparisons of all methods under each example.}
    \label{run_time}
\end{figure}

In terms of efficiency shown in Figure \ref{run_time}, FuzzCoh runs within a few seconds, comparable to the fastest baselines and more than 20 times faster than QCD. Taken together, these results show that FuzzCoh achieves the best trade-off between accuracy, robustness, and runtime efficiency, making it the most reliable method across diverse conditions.

\section{Real EEG data application}
In this section, we evaluate the performance of our method using a real EEG dataset\footnote{\scriptsize This dataset is available at
\url{https://figshare.com/articles/dataset/EEG_driver_drowsiness_dataset/14273687?file=30707285}.}  The dataset contains 2,022 trials of EEG recording collected from 11 individuals during a simulated driving task. Each trial consists of 384 time points (corresponding to 3 seconds at a sampling rate of 128 Hz), with each trial labeled as either drowsy or alert, i.e., \( C = 2 \).\footnote{\scriptsize {The RI for other methods are shwon in Appendix, Section \ref{drowsy_alternative}.}} We emphasize that our method accommodates more than two brain states; however, for the purposes of this application, we fix \( C = 2 \). In our application, we defined four regions that are shown to be related with driving-task \citep{Li2022exploratoryDrivers, talento2024kencoh}, namely, (i) left frontal and pre-frontal lobe (LFp), (ii) right frontal and pre-frontal lobe (RFp), (iii) left temporo-parietal and occipital lobe (LTPO) and (iv) right temporo-parietal and occipital lobe (RTPO) (see leftmost panel of Figure~\ref{framework}). Our goal here is to address the following research questions:

\begin{enumerate}[nosep]
\item What essential insights on brain networks can we uncover, beyond those offered by hard clustering methods, without compromising clustering accuracy?
\item Which frequency band best distinguishes between alert and drowsy states?
\item Which pair of brain regions exhibits the most distinct connectivity between the two brain states?
\end{enumerate}

The last two questions reveal a key limitation of temporal-domain clustering methods. In contrast, our spectral approach identifies discriminative frequency bands and quantifies brain region connectivity, offering deeper insights into spatial interactions. 

Table~\ref{RI_combined_sorted} presents the maximum RI per subject obtained under two settings: (i) using all connectivity vectors (i.e., all four brain regions predefined in Figure~\ref{framework}), within a specified frequency band, and (ii) selecting a pair of brain region within a specific frequency band that yields the highest accuracy per subject. The number of trials for each subject is also reported. The accuracy is evaluated against the labels defined by the experimenters based on the driver's reaction time \citep[see][for details]{Cao2019multi}, which we treat as the ground truth. To have a label that can compare with the true label (i.e., ground truth), we assign the time series to the group using the maximum membership rule. For each configuration, we report the corresponding frequency band, accuracy, and the fuzziness parameter \(m\) that yield the best results, together with the number of fuzziness time series we detected using a 0.7 threshold (see \citep{d2009autocorrelation,lopez2022quantile,ma2025fcpca} for a detailed discussion of the threshold).

\subsection{All subject analysis}
From the frequency-band-based analysis, we observe that the {Beta} and {Gamma} bands frequently achieve the highest clustering accuracy across subjects. Notably, {Beta} is the most dominant, yielding the best performance for 5 out of 11 subjects. This observation is consistent with neurophysiological findings: Beta activity (12–30 Hz) is associated with active thinking, alertness, and focused cognitive processing—states that are typically diminished during drowsiness \citep{sugumar2017eeg}. Similarly, Gamma activity (above 30 Hz) reflects higher-order mental functions, such as attention and sensory integration, which also deteriorate as the brain transitions into drowsy or unconscious states \citep{newson2019eeg}. Therefore, these bands naturally capture discriminative features that separate alert and drowsy states, explaining their superior performance in our clustering framework.

The results that incorporate brain region connectivity provide additional interpretability by linking clustering metrics to underlying neural functions. Subj6 and Subj7 achieve their best accuracy using {Theta} band signals from the LFp--RTPO and LFp--RFp connections, respectively. This pattern highlights the pivotal role of the LFp in distinguishing between alert and drowsy states. The LFp is known to support working memory, executive functions, and language processing—core cognitive domains that are sensitive to fluctuations in mental alertness \citep{beauchamp2005see}. It plays a central role in goal-directed behavior, planning, and error monitoring, and is structurally connected to other regions such as the temporal and occipital cortices. These connections, particularly with RTPO (associated with sensory integration) and RFp (involved in cognitive control), may reflect the breakdown of coordinated top-down regulation as alertness declines. Thus, the prominence of LFp connectivity in achieving high clustering accuracy suggests that changes in executive and attentional networks are key markers of cognitive state transitions.

\begin{table*}[t]  
\centering
\setlength{\tabcolsep}{3pt}
\small   
\centering
\caption{Max RI per subject by frequency band and regional connectivity, with trial counts.}
\label{RI_combined_sorted}
\begin{tabular}{l c c c c c | c c c c c}
\toprule
\small{Subj} & \small{Number of Trials} & \small{RI (\%)} & \small{Band} & $m$ & \small{Fuzzy  series (\%)} &
\small{Connectivity} & \small {RI (\%)} & \small {Band} & $m$ & {Fuzzy series (\%)} \\

\midrule
Subj6  & 166 & 94.57 & Theta & 1.2 & 15.06 & LFp--RTPO  & 85.54 & Theta & 2.2 & 8.43 \\
Subj2  & 132 & 88.64 & Theta & 1.2 & 5.30  & LFp--LTPO  & 89.39 & Beta  & 1.2 & 0.00 \\
Subj9  & 314 & 87.57 & Beta  & 1.8 & 62.42 & LFp--RFp   & 81.84 & Beta  & 2.2 & 16.24 \\
Subj5  & 224 & 86.61 & Beta  & 1.2 & 13.39 & LFp--LTPO  & 79.46 & Beta  & 1.2 & 1.79 \\
Subj11 & 226 & 80.97 & Beta  & 1.8 & 9.73  & RFp--RTPO  & 74.33 & Theta & 1.8 & 1.77 \\
Subj10 & 108 & 75.92 & Gamma & 1.5 & 13.89 & LFp--RFp   & 76.85 & Gamma & 2.2 & 4.62 \\
Subj1  & 188 & 75.53 & Alpha & 1.5 & 33.51 & LFp--RFp   & 78.72 & Beta  & 1.2 & 5.31 \\
Subj4  & 148 & 67.57 & Beta  & 2.5 & 17.57 & LFp--RFp   & 67.57 & Gamma & 1.8 & 18.24 \\
Subj3  & 150 & 67.33 & Beta  & 1.2 & 14.00 & RFp--RTPO  & 70.67 & Gamma & 1.2 & 6.00 \\
Subj8  & 264 & 66.29 & Gamma & 1.2 & 16.67 & LFp--LTPO  & 62.87 & Alpha & 2.0 & 11.36 \\
Subj7  & 102 & 65.69 & Gamma & 1.2 & 21.57 & LFp--RFp   & 73.53 & Theta & 2.2 & 3.92 \\
\bottomrule
\end{tabular}
\end{table*}

\subsection{Selected individual subject analysis}

We further investigate the signals for Subject 2 and Subject 6, which show relatively high RI, i.e., high separation of two states (see Table~\ref{RI_combined_sorted}). Subject 2 has $m=1.2$ fuzziness parameter which indicates that the MTS are well-separable, while Subject 6 has $m=2.2$ fuzziness parameter for LFp-RTPO suggesting presence of fuzzy MTS \footnote{\scriptsize We show in Appendix, Section \ref{auto_select_c_m_drowsy} how to use FSI to guide the selection of hyperparamters in case of no prior infomation of the dataset is known. }.

 \begin{figure}[ht]
    \centering
    \includegraphics[width = 1\linewidth]{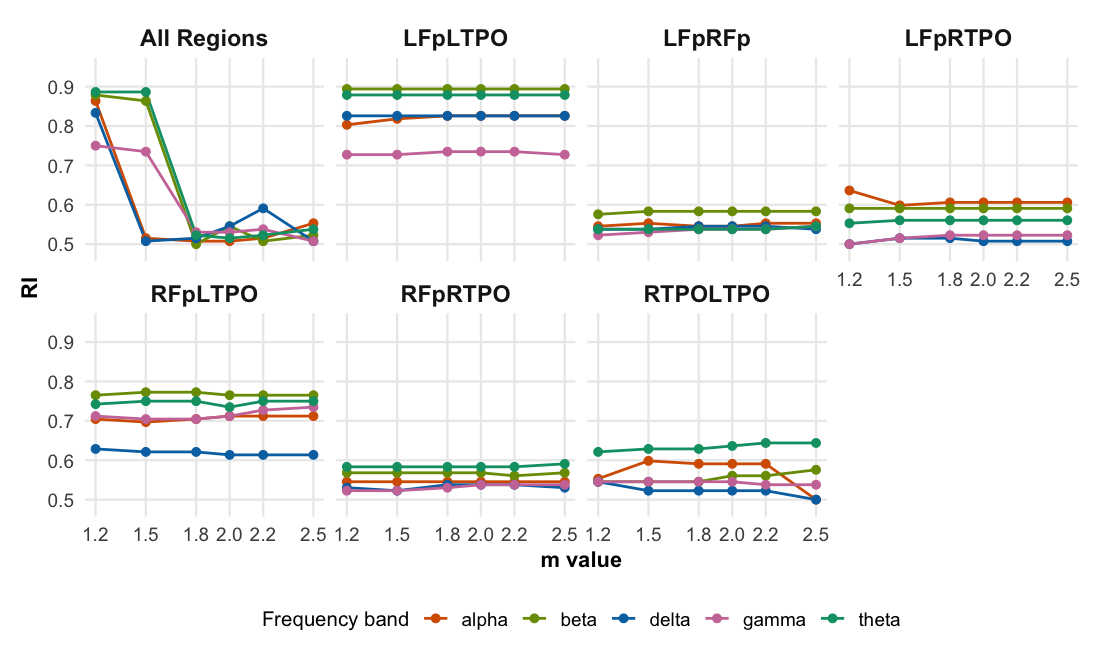}
    \caption{The RI of all or partial brain regions, frequency band, and $m$ value of subject 2.}
    \label{Subject2_reigion}
    \includegraphics[width= 1\linewidth]{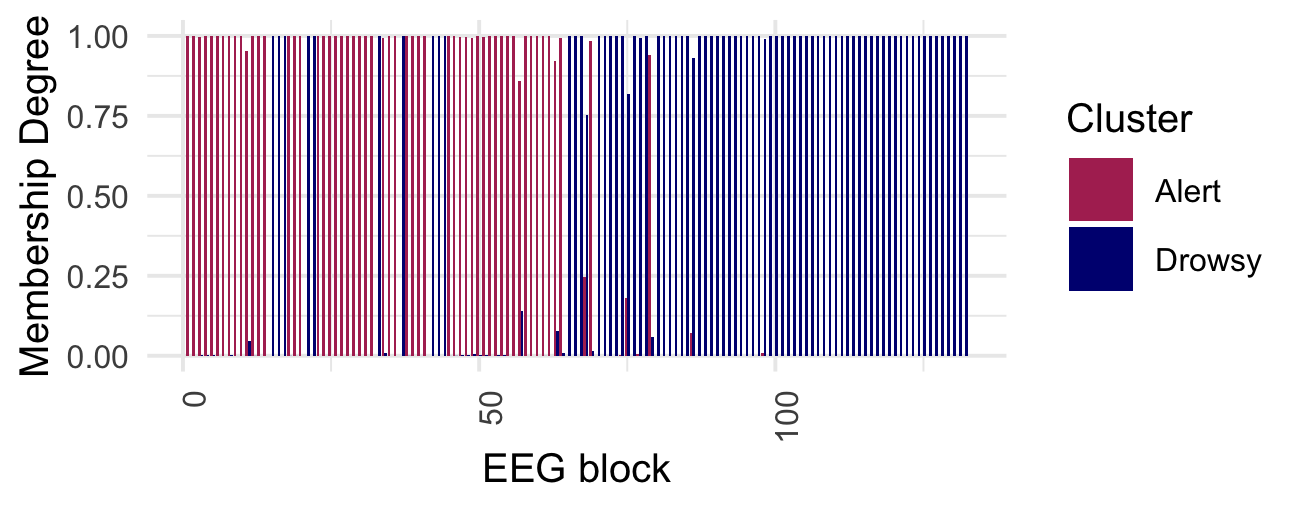}
    \caption{Membership matrix of EEG samples collected from Subject 2 at Beta-band using LFp–LTPO connectivity.}
    \label{subject2_mem}
\end{figure}

    \begin{figure}[ht]
    \centering
    \includegraphics[width=0.6\linewidth]{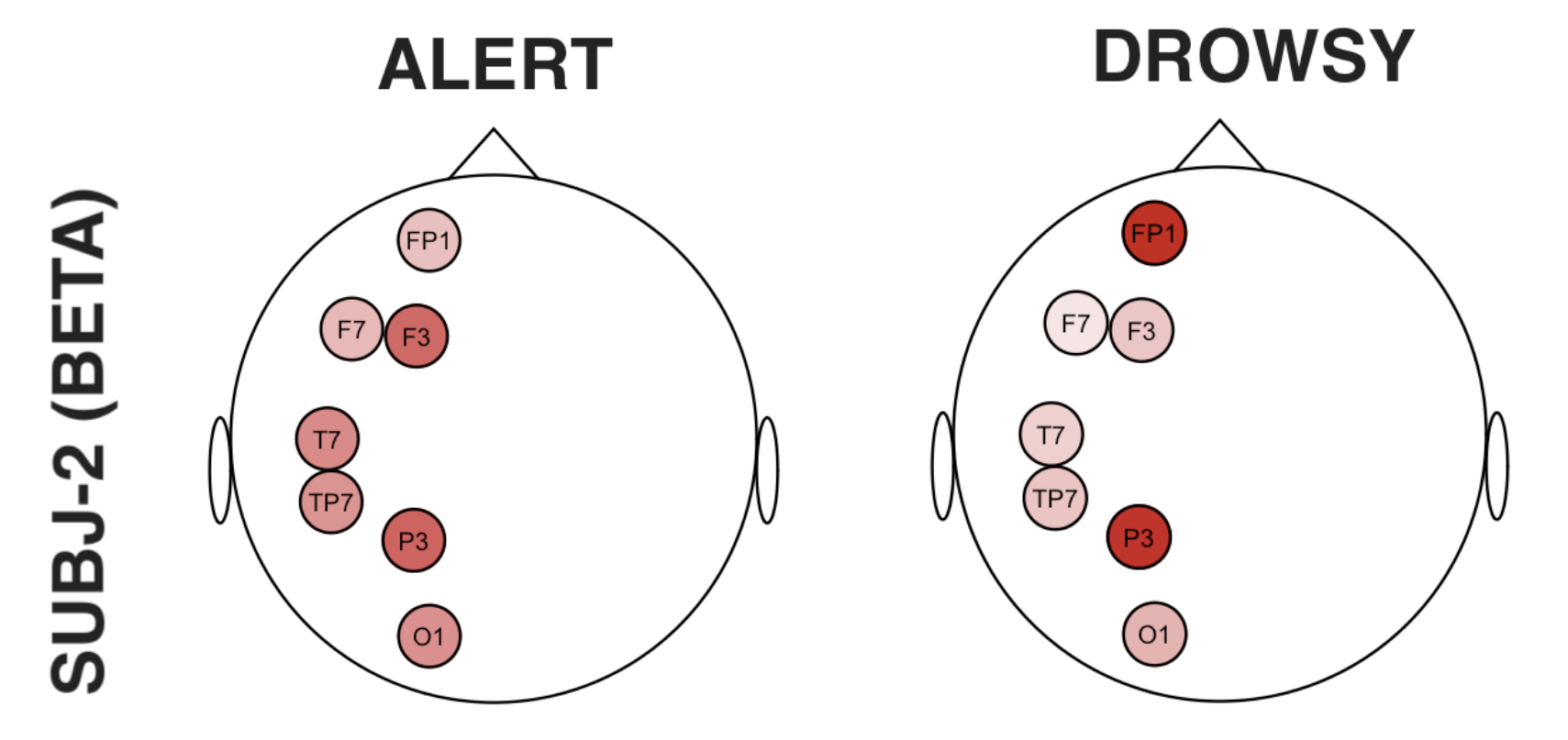}
    \caption{Functional brain connectivity structure (LFp–LTPO) of Subject 2 at Beta band when \textbf{alert} and when \textbf{drowsy}.}
    \label{Subj2_hard}
    \includegraphics[width= 1\linewidth]{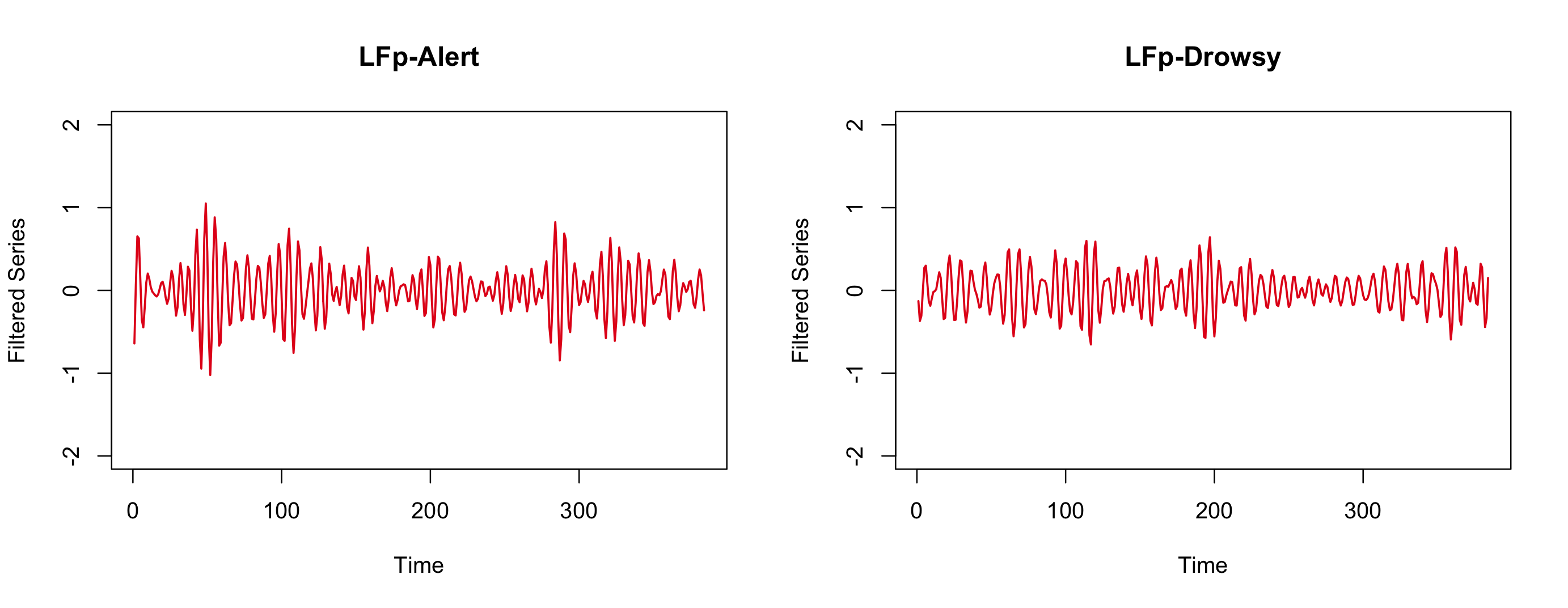}
     \caption{Filtered signals from LTPO region during alert and drowsy states at Beta band of Subject 2.}
     \label{filtered_signal_subject2_LTPO} 
\end{figure} 
\paragraph{Subject 2.} \ \ Figure~\ref{Subject2_reigion} presents the clustering performance of Subject 2 across different brain region pairs and frequency bands, evaluated under varying fuzziness values $m$. The subplot labeled ``All Regions'' shows a clear degradation in clustering quality as $m$ increases, indicating sensitivity to the fuzziness parameter when all connections are used. In contrast, specific region pairs such as LFp--LTPO maintain high and stable RI values across all $m$ values, particularly in the Beta and Theta bands. This robustness highlights that selecting meaningful regional connectivity can significantly enhance clustering consistency. Notably, LFp--LTPO outperforms all other pairs, suggesting it captures strong, stable coherence patterns that distinguish between cognitive states for Subject 2. This underlines the importance of targeted connectivity analysis in improving spectral clustering outcomes. Figure \ref{subject2_mem} plots the membership matrix for each subject 2 EEG trial by specifying $m=1.2$ in the Beta band using LFp--LTPO connectivity. Each vertical bar represents the membership of a trail to two clusters: alert state (blue) and drowsy state (orange). The high bars indicate a strong association with the corresponding state. We can notice that all EEG trails have a high membership in a certain brain state, indicating that the percentage of fuzzy series is 0 \%. 

Figure \ref{Subj2_hard} shows the LFp–LTPO functional connectivity of Subject 2 in the Beta band for the two brain states. Since Table \ref{RI_combined_sorted} suggests that Subject 2 has no fuzzy series when using regional connectivity,  we only show the figure with the discriminating functional brain connectivity. In the alert state (left panel), the connectivity pattern is dominated by frontal and parietal channels F3 and P3, whereas in the drowsy state (right panel), the hub shifts toward the midline and fronto-polar region, with FP1 and P3 becoming the most central nodes.

Figure \ref{filtered_signal_subject2_LTPO} shows how the discriminating brain connectivity in the LTPO region changes between the alert and drowsy states for Subject 2. The top row (multichannel view) shows traces for channels T7, TP7, P3 and O1 are overlaid for the alert (left) and drowsy (right) segments. In the alert epoch, pronounced amplitude bursts appear around times 0 and 200, especially on FP1 and F3. These bursts are markedly weaker in the drowsy epoch, signalling reduced Beta power and synchrony. The bottom row (single-lead summaries) summarizes the filtered signals in each state. The alert trace exhibits higher-amplitude Beta oscillations, whereas the drowsy trace shows a lower-energy, more homogeneous rhythm. Moreover, more channels are contributing to the overall association at the Beta band during the alert state than during the drowsy state. \cite{Kaminski2012beta} found that increased alertness is accompanied by higher EEG activation in the Beta band. 

\begin{figure}[ht]
    \centering
    \includegraphics[width = 1\linewidth]{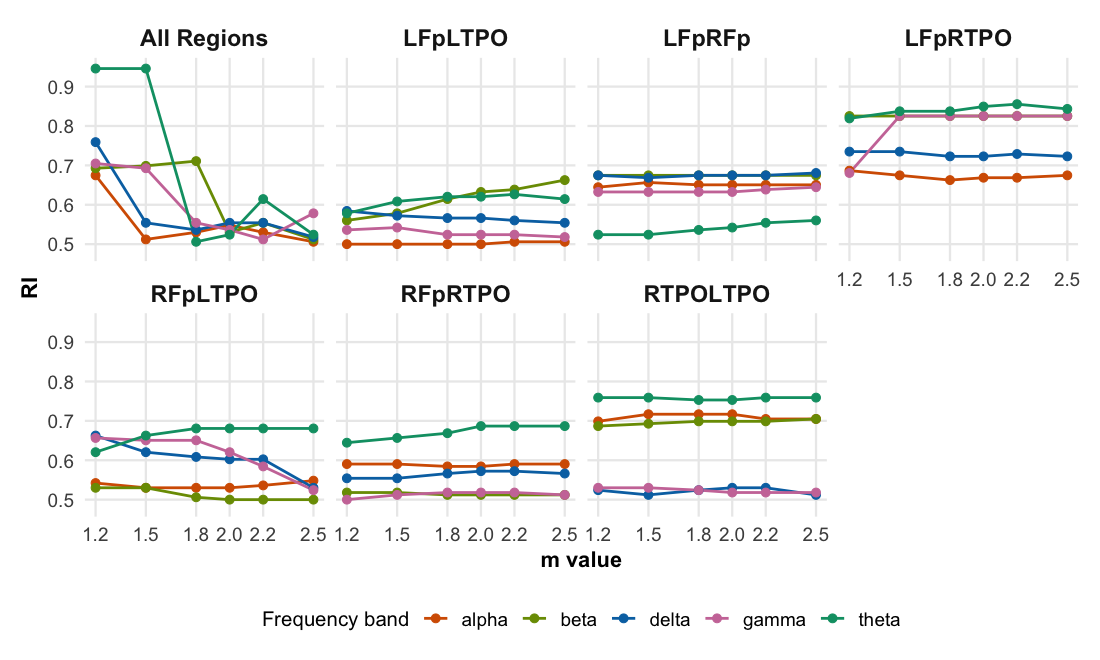}
    \caption{The RI of all or partial brain regions, frequency band, and m value of Subject 6.}
    \label{Subject6_reigion}
    \includegraphics[width=0.8 \textwidth]{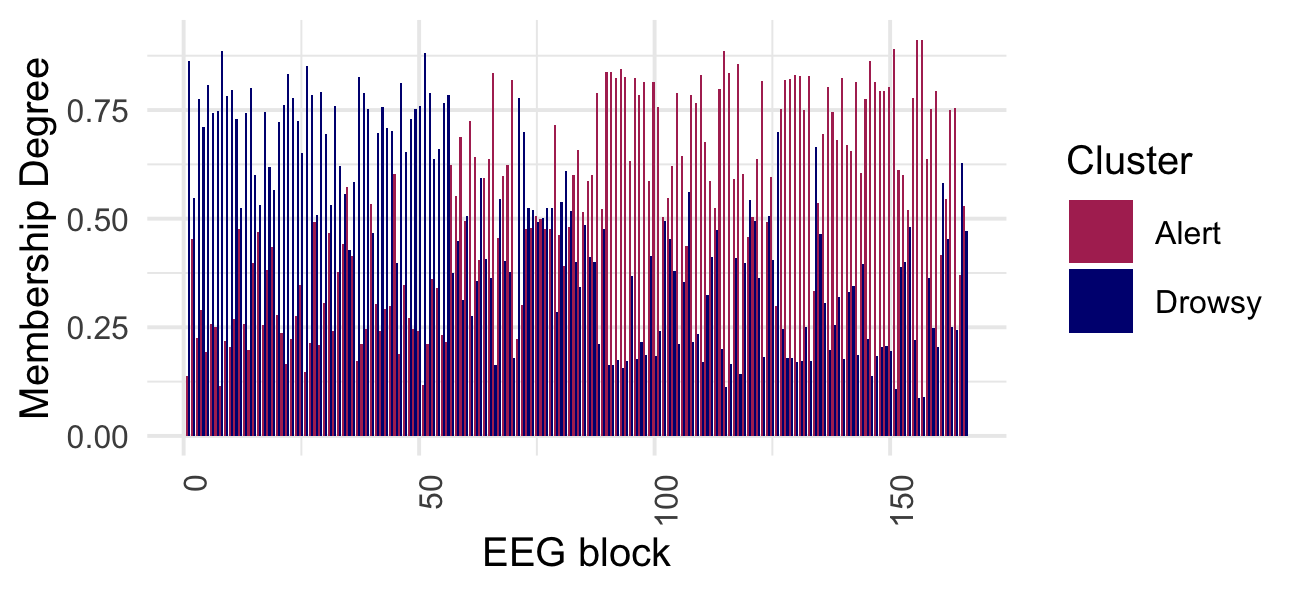}
    \caption{Membership matrix of EEG samples collected from Subject 6 at Theta-band using LFp–RTPO connectivity.}
    \label{subject6_mem}
\end{figure}

\paragraph{Subject 6. } \ \ We show the RI across different brain region pairs, frequency bands, and fuzziness values \(m\) for Subject 6 in detail, as shown in Figure~\ref{Subject6_reigion}. Beta, Theta, and Gamma bands consistently yield higher RI, indicating strong discriminative power for distinguishing alert and drowsy states. Figure~\ref{subject6_mem} presents the membership matrix using Theta-band signals and all brain regions with \(m = 2.2\). Unlike in Figure \ref{subject2_mem}, now some EEG trials exhibit substantial membership in both clusters, indicating the presence of a transitional cognitive state between alertness and drowsiness. This underscores the need to detect and analyze such intermediate stages, which are often missed by hard clustering approaches. This flexibility improves clustering quality by accommodating uncertainty and enabling the detection of subtle shifts in cognitive states.

\begin{figure}[ht]
    \centering
    \includegraphics[width=0.6\textwidth]{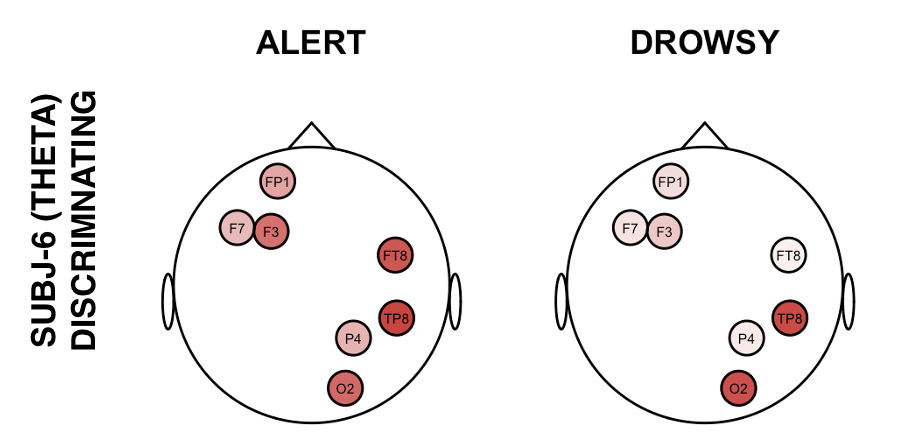}
    \includegraphics[width=0.6\textwidth]{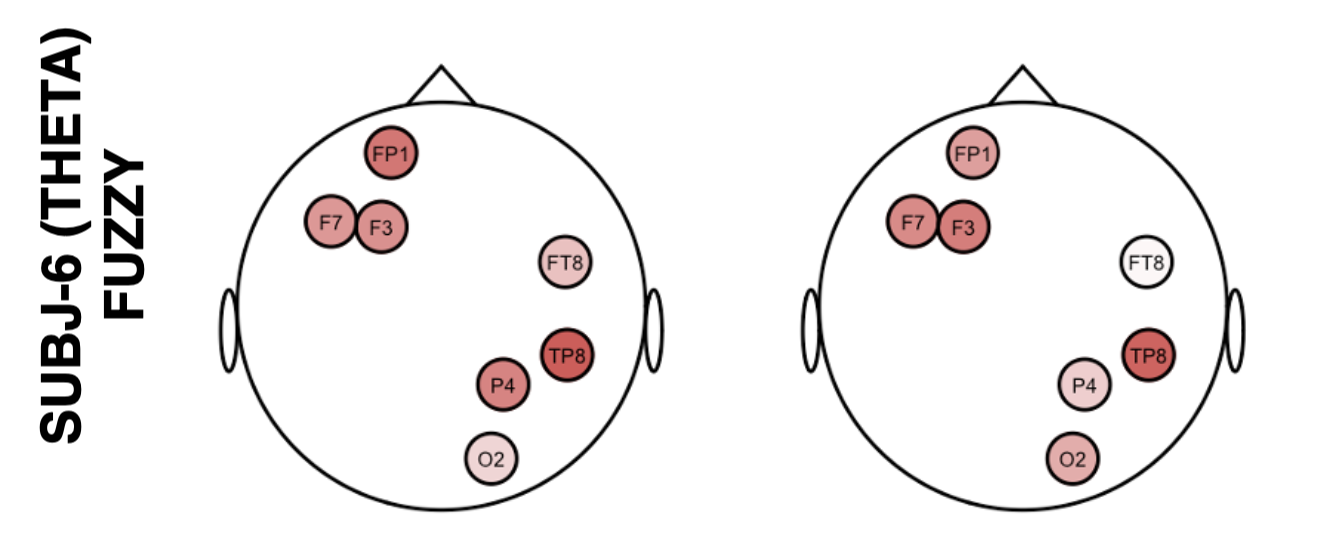}
     \caption{(Top) Discriminating and (bottom) fuzzy functional brain connectivity structure (LFp–RTPO) of Subject 6 at Theta band when \textbf{alert} and when \textbf{drowsy}. }
    \label{Subj6}
  
\end{figure}

\begin{figure}[ht]
    \centering
     \includegraphics[width=0.8\textwidth]{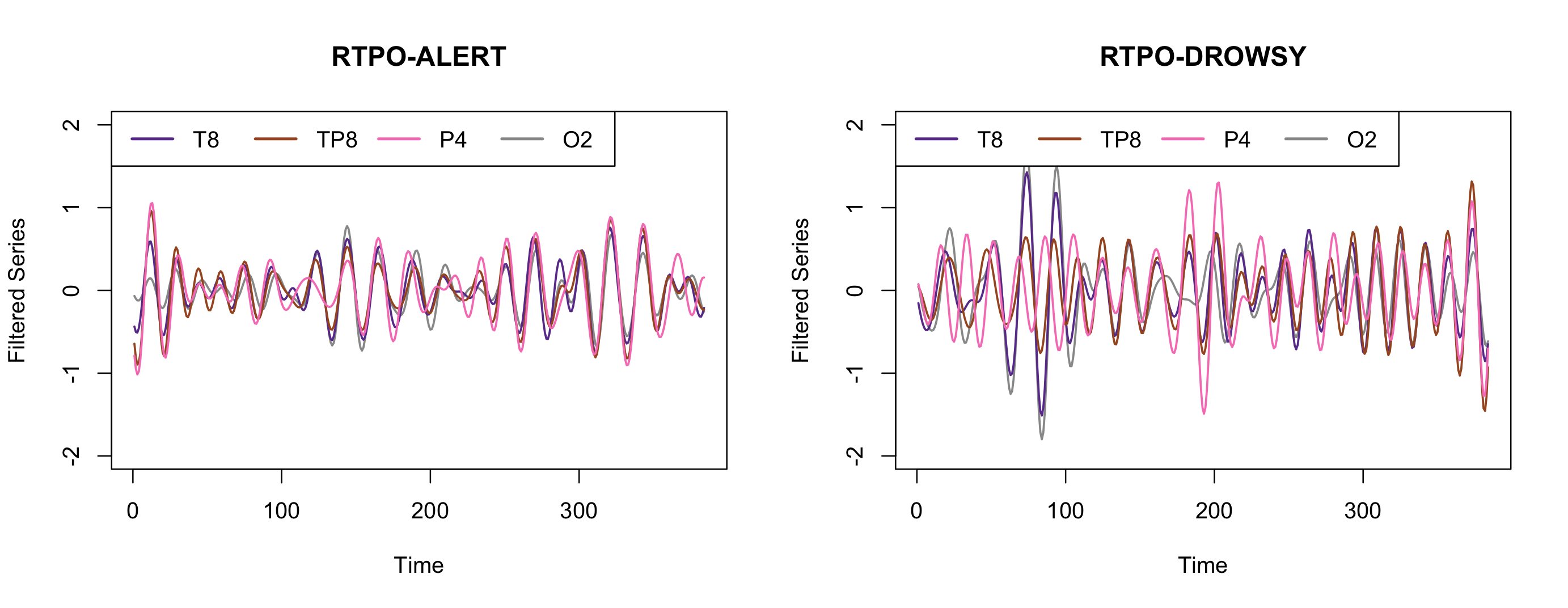}
    \includegraphics[width=0.8\textwidth]{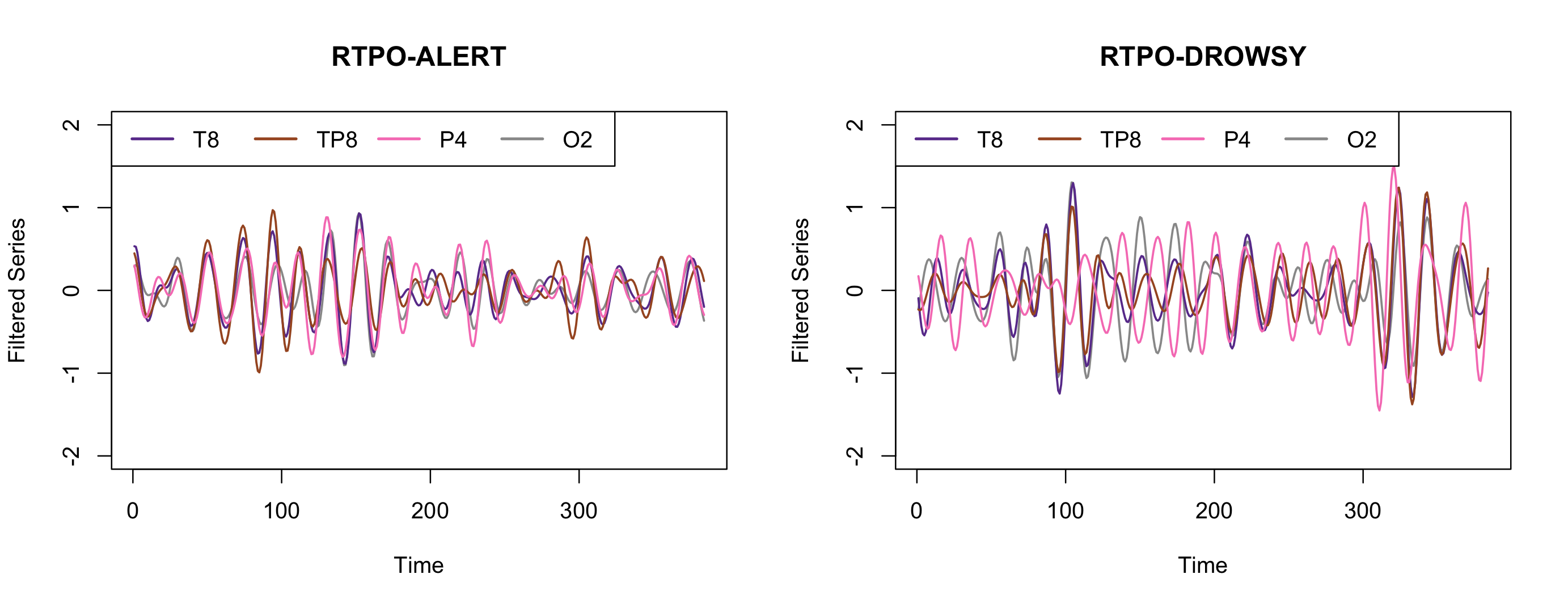}
     \caption{(Top) Discriminating and (bottom) fuzzy filtered signals from RTPO region during alert and drowsy states of Subject 6.}
     \label{discrim_subject6_RTPO}
\end{figure}

As suggested in Table \ref{RI_combined_sorted}, analyzing the regional connectivity, FuzzCoh suggests 8.43\% of its EEG trails are fuzzy series. We would like to visualize both the discriminating and fuzzy scenarios. Figures \ref{Subj6} (top) and \ref{discrim_subject6_RTPO} present the non-fuzzy observations and clearly highlight the distinctions between the two brain states. These observations accord with earlier work: \cite{Brown2012Drowsy} reported a shift toward strong anterior Theta rhythms at the onset of drowsiness, and \cite{Sturm1999TemporalParietal} identified evidence for a fronto-parietal-thalamic-brainstem network in the right hemisphere during alertness of 15 individuals.

The most informative insight comes from the fuzzy cases. Figure \ref{Subj6} (bottom) shows the fuzzy functional connectivity of Subject 6 in the Theta band. Unlike the discriminating trials, the two states display nearly identical connectivity: every node appears in both maps with comparable intensity, and no channel clearly dominates.

Figure \ref{discrim_subject6_RTPO} plot the corresponding filtered time series for RTPO region. The fuzzy waveforms and amplitudes are virtually indistinguishable between the alert (left) and drowsy (right) windows, underscoring the ambiguous nature of these trials. Their connectivity does not align decisively with either cluster, suggesting they represent a transitional or mixed brain state rather than a distinct condition.

Overall, these results suggest that both frequency-specific and connectivity-specific information contribute meaningfully to distinguishing between drowsy and alert brain states. Moreover, tuning the fuzziness parameter \(m\) plays an important role in optimizing clustering performance by balancing cluster separation and membership uncertainty. Importantly, by employing a fuzzy clustering framework, we are also able to capture transitional or mixed cognitive states, where the brain activity may not be fully classified as either drowsy or alert. This capacity to represent ambiguity in cluster membership is particularly valuable for modeling the gradual and continuous nature of cognitive state changes observed in EEG data.

\section{Discussion} 

This paper introduces a robust spectral feature-based fuzzy clustering algorithm for MTS, with EEG data serving as a motivating application. We first outline the concept of canonical variates and then formulate the clustering problem in detail. Through extensive comparisons with alternative methods under both clean and contaminated conditions, we show that our approach consistently achieves superior performance. The main advantages of the proposed FuzzCoh method are: (1) robustness to outliers, (2) graded memberships that capture overlapping cluster structure, (3) operation in the frequency domain with explicit modeling of functional connectivity across multiple locations or variables, and (4) dimension reduction via region-specific connectivity patterns, enabling effective clustering of high-dimensional MTS data using only the most informative features. Beyond EEG, the method is broadly applicable to other spectral–spatial MTS domains.

In terms of practical impact, our approach provides value across diverse fields. In driver drowsiness detection, it enables early-warning systems by identifying subtle changes in spectral and spatial EEG patterns \citep{stancin2021review}. In clinical diagnostics, it can aid the differentiation of neurological disorders through connectivity alterations \citep{smith2005eeg}. For neuromarketing and cognitive workload assessment, it extracts interpretable features reflecting mental effort and engagement \citep{dehais2020neuroergonomics}.

Beyond neuroscience, the FuzzCoh framework extends naturally to other applications. In finance, clustering coherence among asset groups can reveal market regimes and systemic risk \citep{haldane2011systemic,biase2012determinants}, while coherence patterns across cryptocurrencies can be used to detect anomalies and coordinated behavior at multiple frequencies \citep{pocher2023detecting}. In environmental monitoring, it facilitates the clustering of sensor networks (e.g., air quality or temperature) based on shared temporal patterns, enabling early detection of abnormal or synchronized environmental changes \citep{yin2008clustering}.

By modeling both frequency- and region-specific dependencies within a robust fuzzy clustering framework, FuzzCoh serves as a versatile and interpretable tool for analyzing high-dimensional MTS in real-world contexts.

Despite these strengths, FuzzCoh has limitations. Its performance depends on careful parameter tuning, and it assumes local stationarity, an assumption that may not always hold. Moreover, reliance on predefined regions and frequency bands may restrict its adaptability in certain applications. Future work will therefore focus on: (1) developing adaptive block segmentation and scalable approximations for large-scale MTS, (2) extending FuzzCoh to non-stationary settings, and (3) generalizing the framework for robust outlier detection in MTS.

\section*{Acknowledgments}
 This research was supported by King Abdullah University of Science and Technology (KAUST).

\section*{Conflict of interest statement}
The authors declare that they have no known competing financial
interests or personal relationships that could have appeared to
influence the work reported in this paper.

\section*{Data availability statement}
Data sharing is not applicable to this article, as the datasets used in our paper are already publicly available.

\newpage

\section{Appendix for ``Robust Spectral Fuzzy Clustering of Multivariate Time Series with
Applications to Electroencephalogram"}

\subsection{Performance of FuzzCoh beyond EEG using the Cricket dataset}\label{cricket_data}
The Cricket dataset is a MTS benchmark consisting of motion recordings of 12 different umpire signals in cricket.\footnote{\scriptsize{This dataset is available at \url{https://www.timeseriesclassification.com/description.php?Dataset=Cricket}}}. The data were collected using two orthogonal accelerometers, each providing three channels ($X$, $Y$, $Z$ axes), resulting in six-dimensional time series per gesture. Each series has length 1,197, and the dataset contains 72 series in total. For analysis with FuzzCoh, we naturally divide the six channels into two groups corresponding to the two accelerometers. The dataset is challenging because the time series are not strictly aligned across instances, requiring methods that can capture temporal dynamics and discriminate between subtle differences in motion patterns across the 12 gesture classes. In total, 50 replications are conducted for each method.

\subsubsection{On the clean Cricket data }
Table \ref{tab:cricket_clean} shows the RIs of each method. On the clean Cricket data, FuzzCoh achieves the highest RI, outperforming the best time-domain baseline FCPCA by 9\%.  

\begin{table}[ht]
\caption{Comparison of RI on the clean Cricket dataset. The standard deviation is reported in parentheses.}
\centering
\begin{tabular}{c c c c c c}
\toprule
Method & FuzzCoh & FCPCA & QCD & WAV & VPCA \\
\midrule
RI     & 0.41 (0.02) & 0.32 (0.01) & 0.29 (0.09) & 0.27 (0.02) & 0.17 (0.00) \\
\bottomrule
\end{tabular}
\label{tab:cricket_clean}
\end{table}

\subsubsection{On the contaminated Cricket data}
Let $\bs{Z}^{(b)} \in \mathbb{R}^{1197}\times 6$, for $b = 1, \dots, 72$, be the block of time series from test set of Cricket data. Moreover let $\{W_j(t)\}_{t = 1}^{1197} \overset{IID}{\sim} \text{Student's }t_1$ for $j = 1,2,3$. We denote the contaminated three variables in $\bs{Z}^{(b)}$ as $\tilde{Z}_{j}(t) = Z_{j}(t) + 0.1W_j(t)$, for $j \in \{\text{‘X’ in accelerometer 1},\allowbreak\ \text{‘X’ in accelerometer 2},\allowbreak\ \text{‘Z’ in accelerometer 2}\}$.
.
FuzzCoh achieves the highest RI (Table~\ref{tab:cricket_contaminated}), with only a moderate drop compared to the clean case. In contrast, FCPCA and VPCA deteriorate substantially.
\begin{table}[ht]
\caption{Comparison of RI on the contaminated Cricket dataset. The standard deviation is reported in parentheses.}
\centering
\begin{tabular}{c c c c c c}
\toprule
Method & FuzzCoh & FCPCA & QCD & WAV & VPCA \\
\midrule
RI     & 0.35 (0.02) & 0.13 (0.07) & 0.28 (0.01) & 0.26 (0.02) & 0.08 (0.01) \\
\bottomrule
\end{tabular}
\label{tab:cricket_contaminated}
\end{table}

This structure enables us to study not only the within-accelerometer dynamics but also the cross-coherence between the two accelerometers, thereby capturing spatially distributed motion patterns that discriminate between the different umpire signals.

\subsection{The RI using the real EEG data for the alternative methods}\label{drowsy_alternative}
In this section, we present the clustering accuracy of all methods on the real EEG data, as summarized in Table \ref{real_eeg_compare}. For QCD, WAV, and VPCA, we report the RI values obtained using the fuzziness parameter $m$ that yields the highest performance. For FCPCA, we follow the automatic selection of $m$ recommended by its authors. 

FuzzCoh provides the highest and most stable accuracy, with band-based features leading overall and connectivity adding useful, subject-specific gains, supporting the value of combining frequency- and connectivity-level information.

\begin{table}[ht]
\small
\caption{RI on the real EEG data for the comparison methods.}
\centering
\begin{tabular}{c c c c c c c c}
\toprule
Subj & \# Trials & \multicolumn{2}{c}{FuzzCoh} & FCPCA & QCD & WAV & VPCA\\
     &           & By Band & By Conn.                 &       &     &     &     \\
\midrule
1   & 188 & 0.76 & 0.79 & 0.74 & 0.73 & 0.73 & 0.56 \\
2   & 132 & 0.89 & 0.89 & 0.89 & 0.78 & 0.77 & 0.56 \\
3   & 150 & 0.67 & 0.71 & 0.55 & 0.70 & 0.70 & 0.51 \\
4   & 148 & 0.68 & 0.68 & 0.57 & 0.78 & 0.77 & 0.51 \\
5   & 224 & 0.87 & 0.79 & 0.88 & 0.78 & 0.79 & 0.54 \\
6   & 166 & 0.95 & 0.86 & 0.51 & 0.72 & 0.70 & 0.51 \\
7   & 102 & 0.66 & 0.74 & 0.58 & 0.55 & 0.54 & 0.52 \\
8   & 264 & 0.66 & 0.63 & 0.66 & 0.56 & 0.56 & 0.55 \\
9   & 314 & 0.88 & 0.82 & 0.75 & 0.74 & 0.73 & 0.51 \\
10  & 108 & 0.76 & 0.77 & 1.00 & 0.81 & 0.79 & 0.54 \\
11  & 226 & 0.81 & 0.74 & 0.93 & 0.68 & 0.68 & 0.55 \\
\midrule
Mean &     & 0.78 & 0.76 & 0.73 & 0.71 & 0.70 & 0.53 \\
\bottomrule
\end{tabular}
\label{real_eeg_compare}
\end{table}

\subsection{Example of using FSI to perform automatic selection of $C$ and $m$}\label{auto_select_c_m_drowsy}

\begin{figure*}[ht]
  \centering
  \includegraphics[width=0.9\textwidth]{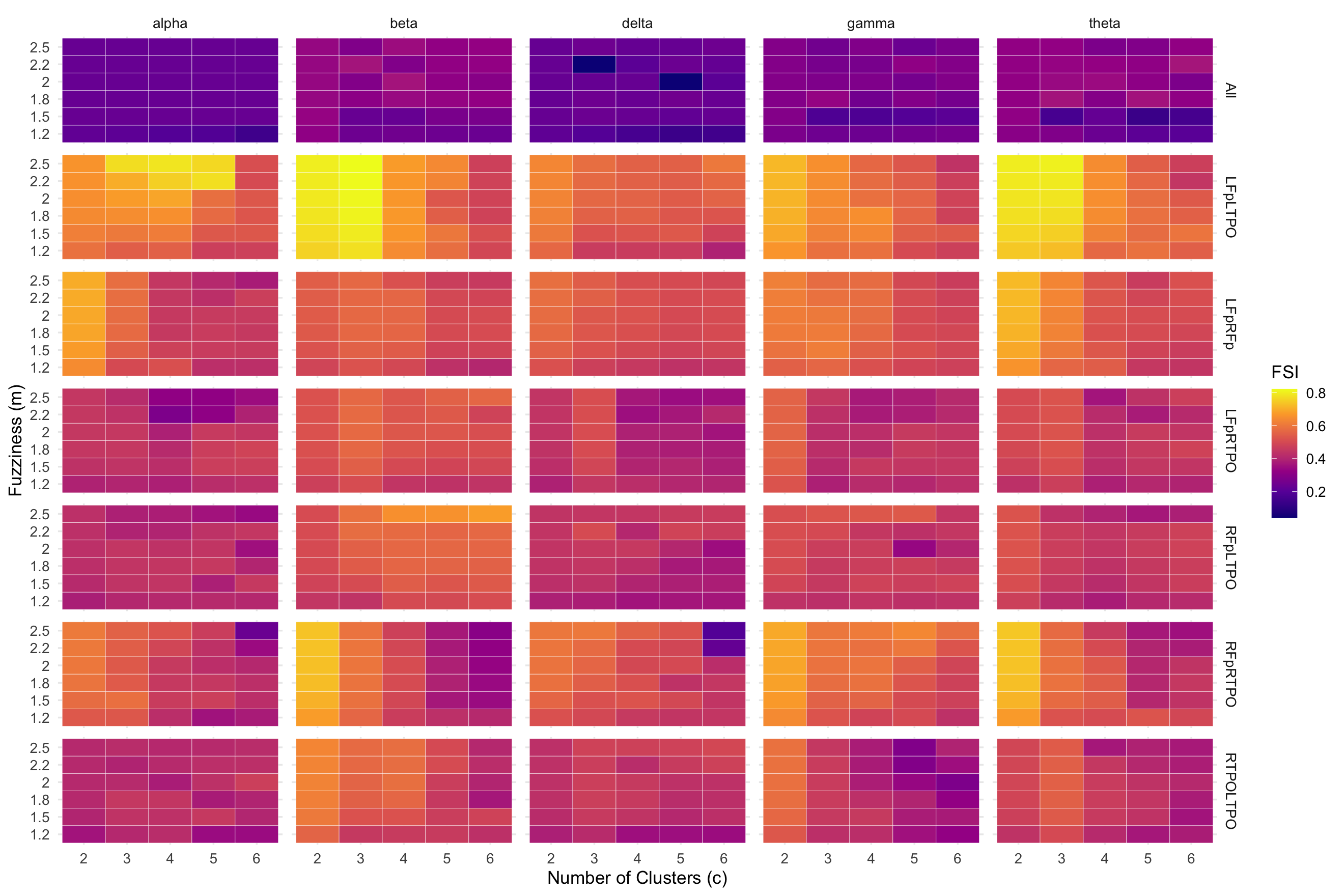}
  \caption{FSI across $(c,m)$ for each region and frequency band of Subject 2.}
  \label{auto_subject2}
\end{figure*}

\begin{figure*}[ht]
      \centering
  \includegraphics[width=0.9\textwidth]{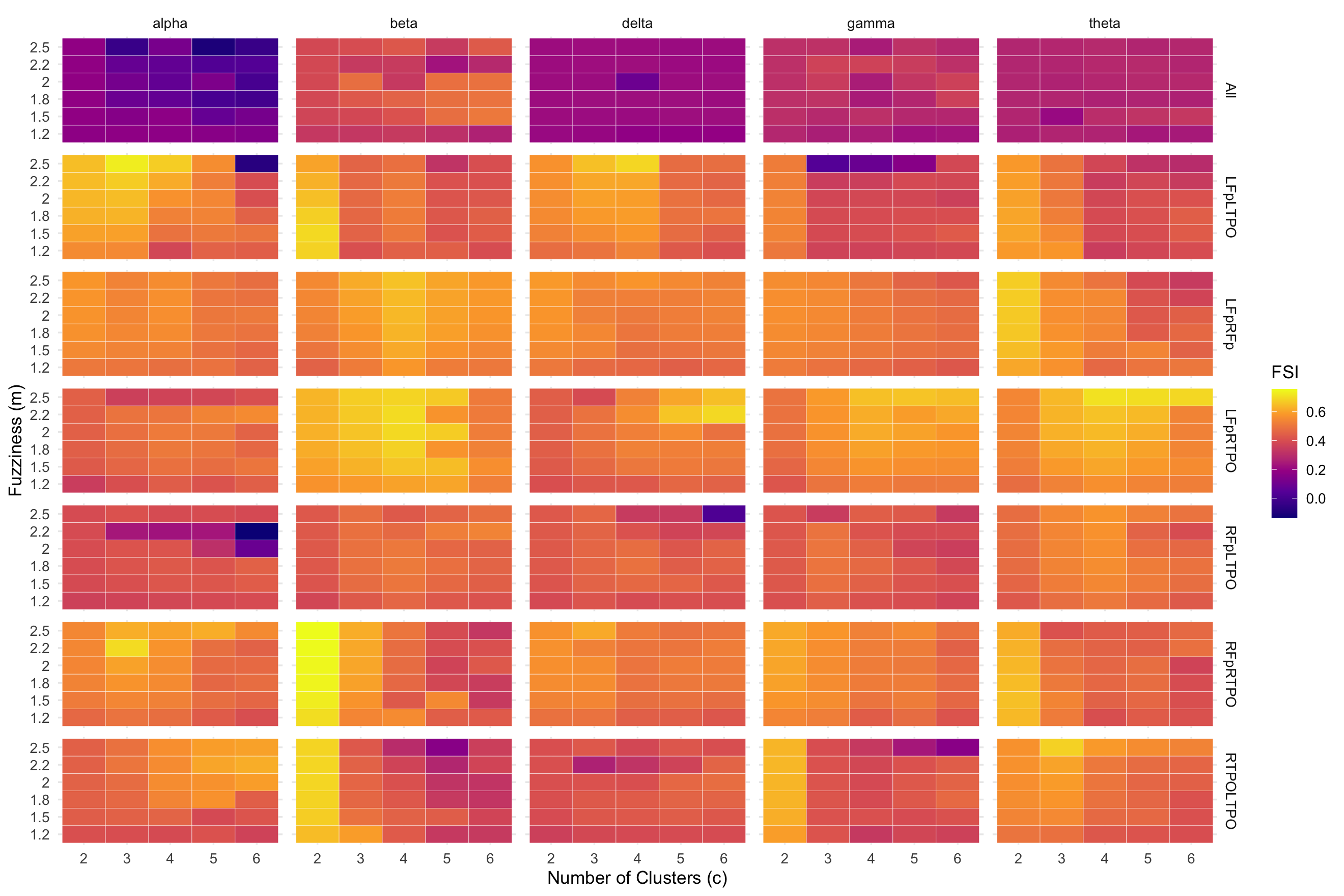}
  \caption{FSI across $(c,m)$ for each region and frequency band of Subject 6.}
  \label{auto_subject6}
\end{figure*}

Figures \ref{auto_subject2} and \ref{auto_subject6}  summarize the FSI over
$C\in\{2,3,\dots,6\}$ and $m\in\{1.2,1.5,1.8,2,2.2,2.5\}$ for each frequency band and inter-regional pair.
For {Subject~2}, the global FSI maximum occurs in Beta at $c=3,m=2.5$ using
LFp--LTPO. For Subject~6, FSI highlights Beta (RFp--RTPO) at $c=2,m=2.5$. The results seem to be different from Table \ref{RI_combined_sorted}. This difference reflects that FSI is unsupervised—favoring compact, well-separated fuzzy clusters, whereas RI measures agreement with external labels. However, due to the only two brain states during collecting the data, we fixed in our application $C=2$ and then compute the clustering accuracy. In reality, when no prior information is known for the data, then the FSI can help guide the optimal selection of the number of clusters. Moreover, accounting for regional connectivity and frequency-specific structure improves partition quality, as evidenced by higher FSI scores.

 \pagebreak  
\clearpage 
\addcontentsline{toc}{chapter}{Bibliography}

\bibliographystyle{biom}
\bibliography{paper}

\begin{thebibliography}{}

\bibitem[\protect\citeauthoryear{Albadawi, Takruri, and Awad}{Albadawi et~al.}{2022}]{albadawi2022review}
Albadawi, Y., Takruri, M., and Awad, M. (2022).
\newblock A review of recent developments in driver drowsiness detection systems.
\newblock {\em Sensors} {\bf 22,} 2069.

\bibitem[\protect\citeauthoryear{Ann~Maharaj, D’Urso, and Galagedera}{Ann~Maharaj et~al.}{2010}]{ann2010wavelet}
Ann~Maharaj, E., D’Urso, P., and Galagedera, D.~U. (2010).
\newblock Wavelet-based fuzzy clustering of time series.
\newblock {\em Journal of Classification} {\bf 27,} 231--275.

\bibitem[\protect\citeauthoryear{Beauchamp}{Beauchamp}{2005}]{beauchamp2005see}
Beauchamp, M.~S. (2005).
\newblock See me, hear me, touch me: multisensory integration in lateral occipital-temporal cortex.
\newblock {\em Current Opinion in Neurobiology} {\bf 15,} 145--153.

\bibitem[\protect\citeauthoryear{Biase and D’Apolito}{Biase and D’Apolito}{2012}]{biase2012determinants}
Biase, P.~d. and D’Apolito, E. (2012).
\newblock The determinants of systematic risk in the italian banking system: A cross-sectional time series analysis.
\newblock {\em International Journal of Economics and Finance} {\bf 4,} 152--164.

\bibitem[\protect\citeauthoryear{Brown, Basheer, McKenna, Strecker, and McCarley}{Brown et~al.}{2012}]{Brown2012Drowsy}
Brown, R.~E., Basheer, R., McKenna, J.~T., Strecker, R.~E., and McCarley, R.~W. (2012).
\newblock Control of sleep and wakefulness.
\newblock {\em Physiological Reviews} {\bf 92,} 1087–1187.

\bibitem[\protect\citeauthoryear{Cao, Chuang, King, and Lin}{Cao et~al.}{2019}]{Cao2019multi}
Cao, Z., Chuang, C.-H., King, J.-K., and Lin, C.-T. (2019).
\newblock Multi-channel {EEG} recordings during a sustained-attention driving task.
\newblock {\em Scientific Data} {\bf 6,} 19.

\bibitem[\protect\citeauthoryear{Dehais, Lafont, Roy, and Fairclough}{Dehais et~al.}{2020}]{dehais2020neuroergonomics}
Dehais, F., Lafont, A., Roy, R., and Fairclough, S. (2020).
\newblock A neuroergonomics approach to mental workload, engagement and human performance.
\newblock {\em Frontiers in Neuroscience} {\bf 14,} 268.

\bibitem[\protect\citeauthoryear{D'Urso and Maharaj}{D'Urso and Maharaj}{2012}]{d2012wavelets}
D'Urso, P. and Maharaj, E.~A. (2012).
\newblock Wavelets-based clustering of multivariate time series.
\newblock {\em Fuzzy Sets and Systems} {\bf 193,} 33--61.

\bibitem[\protect\citeauthoryear{D’Urso, De~Giovanni, and Vitale}{D’Urso et~al.}{2023}]{d2023robust}
D’Urso, P., De~Giovanni, L., and Vitale, V. (2023).
\newblock Robust dtw-based entropy fuzzy clustering of time series.
\newblock {\em Annals of Operations Research} pages 1--35.

\bibitem[\protect\citeauthoryear{D’Urso and Maharaj}{D’Urso and Maharaj}{2009}]{d2009autocorrelation}
D’Urso, P. and Maharaj, E.~A. (2009).
\newblock Autocorrelation-based fuzzy clustering of time series.
\newblock {\em Fuzzy Sets and Systems} {\bf 160,} 3565--3589.

\bibitem[\protect\citeauthoryear{Ferraro, Giordani, Serafini, et~al\mbox{.}}{Ferraro et~al.}{2019}]{ferraro2019fclust}
Ferraro, M.~B., Giordani, P., Serafini, A., et~al. (2019).
\newblock fclust: an {R} package for fuzzy clustering.
\newblock {\em The {R} Journal} {\bf 11,} 1--18.

\bibitem[\protect\citeauthoryear{Filtness, Armstrong, Watson, and Smith}{Filtness et~al.}{2017}]{filtness2017sleep}
Filtness, A.~J., Armstrong, K.~A., Watson, A., and Smith, S.~S. (2017).
\newblock Sleep-related crash characteristics: Implications for applying a fatigue definition to crash reports.
\newblock {\em Accident Analysis \& Prevention} {\bf 99,} 440--444.

\bibitem[\protect\citeauthoryear{Haldane and May}{Haldane and May}{2011}]{haldane2011systemic}
Haldane, A.~G. and May, R.~M. (2011).
\newblock Systemic risk in banking ecosystems.
\newblock {\em Nature} {\bf 469,} 351--355.

\bibitem[\protect\citeauthoryear{He and Tan}{He and Tan}{2018}]{he2018unsupervised}
He, H. and Tan, Y. (2018).
\newblock Unsupervised classification of multivariate time series using {VPCA} and fuzzy clustering with spatial weighted matrix distance.
\newblock {\em IEEE Transactions on Cybernetics} {\bf 50,} 1096--1105.

\bibitem[\protect\citeauthoryear{Higgins, Michael, Austin, {\AA}kerstedt, Van~Dongen, Watson, Czeisler, Pack, and Rosekind}{Higgins et~al.}{2017}]{higgins2017asleep}
Higgins, J.~S., Michael, J., Austin, R., {\AA}kerstedt, T., Van~Dongen, H.~P., Watson, N., Czeisler, C., Pack, A.~I., and Rosekind, M.~R. (2017).
\newblock Asleep at the wheel—the road to addressing drowsy driving.
\newblock {\em Sleep} {\bf 40,} zsx001.

\bibitem[\protect\citeauthoryear{Honey, K{\"o}tter, Breakspear, and Sporns}{Honey et~al.}{2007}]{honey2007network}
Honey, C.~J., K{\"o}tter, R., Breakspear, M., and Sporns, O. (2007).
\newblock Network structure of cerebral cortex shapes functional connectivity on multiple time scales.
\newblock {\em Proceedings of the National Academy of Sciences} {\bf 104,} 10240--10245.

\bibitem[\protect\citeauthoryear{Izakian, Pedrycz, and Jamal}{Izakian et~al.}{2015}]{izakian2015fuzzy}
Izakian, H., Pedrycz, W., and Jamal, I. (2015).
\newblock Fuzzy clustering of time series data using dynamic time warping distance.
\newblock {\em Engineering Applications of Artificial Intelligence} {\bf 39,} 235--244.

\bibitem[\protect\citeauthoryear{Jiang, Bian, and Tian}{Jiang et~al.}{2019}]{jiang2019removalartifacts}
Jiang, X., Bian, G.-B., and Tian, Z. (2019).
\newblock Removal of artifacts from {EEG} signals: a review.
\newblock {\em Sensors} {\bf 19,} 987.

\bibitem[\protect\citeauthoryear{Kami{\'n}ski, Brzezicka, Gola, and Wr{\'o}bel}{Kami{\'n}ski et~al.}{2012}]{Kaminski2012beta}
Kami{\'n}ski, J., Brzezicka, A., Gola, M., and Wr{\'o}bel, A. (2012).
\newblock Beta band oscillations engagement in human alertness process.
\newblock {\em International Journal of Psychophysiology} {\bf 85,} 125--128.

\bibitem[\protect\citeauthoryear{Li, Yang, and Yan}{Li et~al.}{2022}]{Li2022exploratoryDrivers}
Li, X., Yang, L., and Yan, X. (2022).
\newblock An exploratory study of drivers’ {EEG} response during emergent collision avoidance.
\newblock {\em Journal of Safety Research} {\bf 82,} 241--250.

\bibitem[\protect\citeauthoryear{Little and Brown}{Little and Brown}{2014}]{Little2014BetaParkinson}
Little, S. and Brown, P. (2014).
\newblock The functional role of beta oscillations in parkinson's disease.
\newblock {\em Parkinsonism \& Related Disorders} {\bf 20,} S44--S48.

\bibitem[\protect\citeauthoryear{L{\'o}pez-Oriona, D'Urso, Vilar, and Lafuente-Rego}{L{\'o}pez-Oriona et~al.}{2022}]{lopez2022quantile}
L{\'o}pez-Oriona, {\'A}., D'Urso, P., Vilar, J.~A., and Lafuente-Rego, B. (2022).
\newblock Quantile-based fuzzy c-means clustering of multivariate time series: Robust techniques.
\newblock {\em International Journal of Approximate Reasoning} {\bf 150,} 55--82.

\bibitem[\protect\citeauthoryear{Ma, Ángel López-Oriona, Ombao, and Sun}{Ma et~al.}{2025}]{ma2025fcpca}
Ma, Z., Ángel López-Oriona, Ombao, H., and Sun, Y. (2025).
\newblock {FCPCA: Fuzzy clustering of high-dimensional time series based on common principal component analysis}.
\newblock {\em International Journal of Approximate Reasoning} {\bf 187,} 109552.

\bibitem[\protect\citeauthoryear{Maharaj and D’Urso}{Maharaj and D’Urso}{2011}]{maharaj2011fuzzy}
Maharaj, E.~A. and D’Urso, P. (2011).
\newblock Fuzzy clustering of time series in the frequency domain.
\newblock {\em Information Sciences} {\bf 181,} 1187--1211.

\bibitem[\protect\citeauthoryear{Masulli, Masulli, Rovetta, Lintas, and Villa}{Masulli et~al.}{2019}]{masulli2019fuzzy}
Masulli, P., Masulli, F., Rovetta, S., Lintas, A., and Villa, A.~E. (2019).
\newblock Fuzzy clustering for exploratory analysis of eeg event-related potentials.
\newblock {\em IEEE Transactions on Fuzzy Systems} {\bf 28,} 28--38.

\bibitem[\protect\citeauthoryear{Newson and Thiagarajan}{Newson and Thiagarajan}{2019}]{newson2019eeg}
Newson, J.~J. and Thiagarajan, T.~C. (2019).
\newblock {EEG} frequency bands in psychiatric disorders: a review of resting state studies.
\newblock {\em Frontiers in Human Neuroscience} {\bf 12,} 521.

\bibitem[\protect\citeauthoryear{Ombao and Pinto}{Ombao and Pinto}{2024}]{ombao2024spectral}
Ombao, H. and Pinto, M. (2024).
\newblock Spectral dependence.
\newblock {\em Econometrics and Statistics} {\bf 32,} 122--159.

\bibitem[\protect\citeauthoryear{Pocher, Zichichi, Merizzi, Shafiq, and Ferretti}{Pocher et~al.}{2023}]{pocher2023detecting}
Pocher, N., Zichichi, M., Merizzi, F., Shafiq, M.~Z., and Ferretti, S. (2023).
\newblock Detecting anomalous cryptocurrency transactions: An {AML/CFT} application of machine learning-based forensics.
\newblock {\em Electronic Markets} {\bf 33,} 37.

\bibitem[\protect\citeauthoryear{Rawashdeh and Ralescu}{Rawashdeh and Ralescu}{2012}]{rawashdeh2012crisp}
Rawashdeh, M. and Ralescu, A. (2012).
\newblock Crisp and fuzzy cluster validity: Generalized intra-inter silhouette index.
\newblock In {\em 2012 Annual Meeting of the North American Fuzzy Information Processing Society (NAFIPS)}, pages 1--6. IEEE.

\bibitem[\protect\citeauthoryear{Rhee and Oh}{Rhee and Oh}{1996}]{rhee1996validity}
Rhee, H.-S. and Oh, K.-W. (1996).
\newblock A validity measure for fuzzy clustering and its use in selecting optimal number of clusters.
\newblock In {\em Proceedings of IEEE 5th International Fuzzy Systems}, volume~2, pages 1020--1025. IEEE.

\bibitem[\protect\citeauthoryear{Rodrigues, Belo, and Gamboa}{Rodrigues et~al.}{2017}]{rodrigues2017noise}
Rodrigues, J., Belo, D., and Gamboa, H. (2017).
\newblock Noise detection on {ECG} based on agglomerative clustering of morphological features.
\newblock {\em Computers in Biology and Medicine} {\bf 87,} 322--334.

\bibitem[\protect\citeauthoryear{Rousseeuw}{Rousseeuw}{1987}]{rousseeuw1987silhouettes}
Rousseeuw, P.~J. (1987).
\newblock Silhouettes: a graphical aid to the interpretation and validation of cluster analysis.
\newblock {\em Journal of Computational and Applied Mathematics} {\bf 20,} 53--65.

\bibitem[\protect\citeauthoryear{Smith}{Smith}{2005}]{smith2005eeg}
Smith, S.~J. (2005).
\newblock {EEG} in the diagnosis, classification, and management of patients with epilepsy.
\newblock {\em Journal of Neurology, Neurosurgery \& Psychiatry} {\bf 76,} ii2--ii7.

\bibitem[\protect\citeauthoryear{Stancin, Cifrek, and Jovic}{Stancin et~al.}{2021}]{stancin2021review}
Stancin, I., Cifrek, M., and Jovic, A. (2021).
\newblock A review of {EEG} signal features and their application in driver drowsiness detection systems.
\newblock {\em Sensors} {\bf 21,} 3786.

\bibitem[\protect\citeauthoryear{Sturm, De~Simone, Krause, Specht, Hesselmann, Radermacher, Herzog, Tellmann, M{\"u}ller-G{\"a}rtner, and Willmes}{Sturm et~al.}{1999}]{Sturm1999TemporalParietal}
Sturm, W., De~Simone, A., Krause, B., Specht, K., Hesselmann, V., Radermacher, I., Herzog, H., Tellmann, L., M{\"u}ller-G{\"a}rtner, H.-W., and Willmes, K. (1999).
\newblock Functional anatomy of intrinsic alertness: evidence for a fronto-parietal-thalamic-brainstem network in the right hemisphere.
\newblock {\em Neuropsychologia} {\bf 37,} 797--805.

\bibitem[\protect\citeauthoryear{Sugumar and Vanathi}{Sugumar and Vanathi}{2017}]{sugumar2017eeg}
Sugumar, D. and Vanathi, P. (2017).
\newblock {EEG} signal separation using improved eemd-fast iva algorithm.
\newblock {\em Asian Journal of Research in Social Sciences and Humanities} {\bf 7,} 1230--1243.

\bibitem[\protect\citeauthoryear{Talento, Roy, and Ombao}{Talento et~al.}{2025}]{talento2024kencoh}
Talento, M. S.~D., Roy, S., and Ombao, H.~C. (2025).
\newblock Ken{C}oh: A ranked-based canonical coherence.
\newblock {\em arXiv preprint arXiv:2412.10521} .

\bibitem[\protect\citeauthoryear{Tefft}{Tefft}{2010}]{tefft2010asleep}
Tefft, B.~C. (2010).
\newblock Asleep at the wheel: The prevalence and impact of drowsy driving.
\newblock {\em AAA Foundation for Traffic Safety} .

\bibitem[\protect\citeauthoryear{Van Den~Heuvel and Pol}{Van Den~Heuvel and Pol}{2010}]{van2010exploring}
Van Den~Heuvel, M.~P. and Pol, H. E.~H. (2010).
\newblock Exploring the brain network: a review on resting-state f{MRI} functional connectivity.
\newblock {\em European Neuropsychopharmacology} {\bf 20,} 519--534.

\bibitem[\protect\citeauthoryear{Van~Vugt, Sederberg, and Kahana}{Van~Vugt et~al.}{2007}]{van2007comparison}
Van~Vugt, M.~K., Sederberg, P.~B., and Kahana, M.~J. (2007).
\newblock Comparison of spectral analysis methods for characterizing brain oscillations.
\newblock {\em Journal of Neuroscience Methods} {\bf 162,} 49--63.

\bibitem[\protect\citeauthoryear{Von~Luxburg}{Von~Luxburg}{2007}]{von2007tutorial}
Von~Luxburg, U. (2007).
\newblock A tutorial on spectral clustering.
\newblock {\em Statistics and Computing} {\bf 17,} 395--416.

\bibitem[\protect\citeauthoryear{Wu, Knight, Cooper, Fortin, and Ombao}{Wu et~al.}{2025}]{wu2025wavelet}
Wu, H., Knight, M.~I., Cooper, K., Fortin, N.~J., and Ombao, H. (2025).
\newblock Wavelet canonical coherence for nonstationary signals: {NeurIPS} 2025 (spotlight).
\newblock In {\em NeurIPS 2025}.

\bibitem[\protect\citeauthoryear{Yin and Gaber}{Yin and Gaber}{2008}]{yin2008clustering}
Yin, J. and Gaber, M.~M. (2008).
\newblock Clustering distributed time series in sensor networks.
\newblock In {\em 2008 Eighth IEEE International Conference on Data Mining}, pages 678--687. IEEE.

\bibitem[\protect\citeauthoryear{Zhao, Cook, Murray, Kesan, Belacel, Doesburg, Medvedev, Vakorin, and Xi}{Zhao et~al.}{2024}]{zhao2024leveraging}
Zhao, C., Cook, Z., Murray, L., Kesan, J., Belacel, N., Doesburg, S., Medvedev, G., Vakorin, V., and Xi, P. (2024).
\newblock Leveraging large language models and fuzzy clustering for eeg report analysis.
\newblock In {\em 2024 IEEE SENSORS}, pages 1--4. IEEE.

\end{thebibliography}

\end{document}